\documentclass[12pt]{article}
\usepackage{epsfig}
      \def\di{\displaystyle}
      \def\br{{\bf r}}
      \def\bp{{\bf p}}
      
      \def\bs{{\bf s}}
      \def\R{{\cal R}}
      \def\P{{\cal P}}

      \def\L{{\cal L}}
      \def\J{{\cal J}}
      \def\Y{{\cal Y}}
      \def\Z{{\cal Z}}
\begin{document}

\vskip 5mm
\begin{center}
{\large\bf THE NUCLEAR SCISSORS MODE IN A SOLVABLE MODEL}\\
\vspace*{1cm}
{\large E.B. Balbutsev}\\
\vspace*{0.2cm}
{\it Joint Institute for Nuclear Research, 141980 Dubna, Moscow Region,
Russia}\\
\vspace*{0.5cm}
{\large P. Schuck}\\
\vspace*{0.2cm}
{\it Institut de Physique Nucleaire, Orsay Cedex 91406, France}
\end{center}

      \vspace{3cm}

\begin{abstract}
 The coupled dynamics of the scissors mode and the isovector giant
quadrupole resonance is studied in a model with separable
quadrupole-quadrupole residual interactions. The method of Wigner
function moments
is applied to derive the dynamical equations for angular momentum and
quadrupole moment. Analytical expressions for energies, B(M1)- and
B(E2)-values, sum rules and flow-patterns of both modes are found
for arbitrary values of the deformation parameter. Some predictions
for the case of superdeformation are given. The subtle nature of the
phenomenon and its peculiarities are clarified.
\end{abstract}

\newpage

\section{Introduction}
The low-energy orbital magnetic-dipole excitations of deformed
nuclei, commonly called the scissors mode, were predicted about 25
years ago \cite{Hilt,Lo}. The prediction was inspired by the
geometric picture that the (prolate) deformed neutron and proton
distributions counter rotate (like scissors) within a small opening
angle $\delta\phi$ in an oscillatory way around their common axis
perpendicular to the long symmetry axis of the nucleus.
Only a few years later the idea was confirmed experimentally with its
detection in $^{156}$Gd \cite{Bohle}. At present, in addition to the
rare earth nuclei, this excitation is also known in the actinides and
in light nuclei. A complete review of the experimental situation can
be found in \cite{Richter}. The discovery of the scissors mode has
initiated a cascade of theoretical studies. An excellent review of the
present situation in this field is the one by D. Zawischa \cite{Zaw}
(see also \cite{Lo2000}).
Very briefly the situation can be described in the following way. All
microscopic calculations with effective forces reproduce
experimental data with respect to the position and the strength of the
scissors mode, some of them \cite{Sushkov}
giving also reasonable fragmentation of its
strength. However, the situation is more obscure in regard to simple
phenomenological models whose aim is to explain the physics of the
phenomenon and to interpret it in the most simple
and transparent terms. A noticeable discord of the opinions  of
various authors must be observed here as has been pointed out in
\cite{Zaw}. One is forced to conclude that there is no
general agreement in the understanding of the nature of this curious
phenomenon. We here will try to shed some light into the confusing
situation.

 Our model Hamiltonian will be the one of the well served harmonic
oscillator plus separable Quadrupole-Quadrupole (Q-Q)
residual interaction.
The interaction will have isoscalar and isovector parts whose coupling
constants are reasonably well known from the literature. Of course,
given such a simple schematic Hamiltonian the RPA equations, which are
the standard tool to describe the scissors mode, can readily be solved.
This, however, does not advance us in its physical interpretation.
Namely, numerous calculations, performed during 25 years of
investigation of the scissors mode have undoubtedly demonstrated
at least one fact: the scissors mode is a very
non-trivial, subtle type of motion subject to the influence of many
factors. In fact its nature is much more complicated and interesting
than the above picture of two counter rotating and oscillating pieces
of matter might suggest. First of all one easily imagines that the
rotational oscillations of neutron and proton systems are inevitably
accompanied by the (quadrupole) distortion of their shapes. This
entails that the scissors mode is intricately entangled with the
Iso-Vector Giant Quadrupole Resonance (IVGQR). Thus, if one wants to
observe the scissors mode, one has to excite at the same time the
IVGQR.
 The quadrupole distortions give rise to high lying excitations
because of the so-called nuclear elasticity (or Fermi surface
deformation), the quantum effect discovered by Bertsch \cite{Bertsch}.
It turns out that
without this additional restoring force the scissors mode would
actually be a zero energy mode! Hence, the scissors mode is a pure
quantum mechanical phenomenon, which can not be explained in the frame
of classical mechanics. The above properties of the scissors mode
highlight
perhaps its most characteristic features. They will obtain a natural,
sufficiently simple, and visual explanation in the frame of our
approach outlined below.

One further issue, strongly debated in the literature, is whether the
neutron and proton fluids really spatially separate at least to a
certain extent during the scissors motion. With diffuse surfaces and
small amplitude motion this is a not completely trivial question and
it only makes sense to speak about the separation of the symmetry axes
of neutron and proton distributions. The debate is not without
foundation, since there were attempts \cite{Zaw} to construct a model
with sharp surfaces: neutron and proton liquids perform out-of-phase
rotational oscillations with Steinwedel-Jensen
boundary conditions. Our approach allows us to derive the analytic
expressions for the current lines and the corresponding figures show
unambiguously that indeed a separation of the two fluids occurs. More
surprisingly, the separation not only exists in the low energy magnetic
dipole excitation $1^+$ (the scissors mode proper) but also
in the $K^{\pi}=1^+$ branch of the IVGQR, which is named "the high
energy scissors mode" and whose existence was guessed by
various authors many years ago \cite{Hilt92,Lo2000}.

 From the above discussion it seems clear that in order to elucidate
all these subtle features an approach involving macroscopic quantities
as dynamical variables is indicated. Most naturally this can be
achieved by working in phase space
\cite{Kohl,Bal}. This can be easily performed by
applying the Wigner transform to the Time Dependent Hartree-Fock (TDHF)
equations. The method of the Wigner Function Moments (WFM)
\cite{Bal,BaSc} is then applied.
It can be characterized as a link between microscopic
and macroscopic approaches: starting from a microscopic Hamiltonian
one derives macroscopic dynamic equations for collective
variables.
 Usually one has to establish the set of such variables with the help
of some physical considerations. The WFM method allows one to avoid
this non-trivial problem: if one knows at least one collective
variable (in our case it is the relative angular momentum of neutrons
and protons), the procedure of derivation of dynamical equations will
automatically generate all the other variables needed.

In this way an unambiguous set of coupled equations in terms of dynamic
physical variables is obtained, which in the small amplitude limit
(RPA) allows for analytic solutions. Since our equations are written
down in the laboratory frame the total angular momentum $I$ is, of
course, conserved. In this work we study the case without rotation and
take $I=0$. These remarks are important, since microscopic calculations
\cite{Zaw} have shown that for the results to be reasonable, it is very
important to exclude from the wave function the spurious component
responsible for the rotation of the nucleus, as a whole. Such problems
do not arise in our approach, because there is no necessity to introduce
the intrinsic coordinate system.

The analytical form of our results is very convenient to study the
deformation ($\delta$) dependence of various quantities such as
position of resonances and transitions probabilities. In the small
$\delta$ limit we mostly reproduce results already obtained by other
authors. However, for large $\delta$ we obtain predictions for super-
and hyper-deformed nuclei. This area is practically not investigated
at present. The only investigation within a phenomenological model
\cite{Lo2000} and the only existing microscopic calculation
\cite{Hamam} are in rather good agreement with our results.
In \cite{Hamam} pairing is taken
into account whereas we have, so far, not considered superfluidity in
our approach. However, it is well known that at large deformations
superfluidity is of little influence on the dynamics and this is also
the conclusion in \cite{Hamam}.
On the other hand at small and moderate $\delta$ the
influence of pairing may be appreciable and certainly our approach
must be generalized to include superfluidity in the future. It is for
example shown that pairing reduces B(M1) - transition probabilities
by important factors and that this yields agreement with the
experimental deformation dependence \cite{Zaw}. In spite of
the importance of pairing correlations in nuclei we know from other
features that it certainly changes results quantitatively but it is
too weak in nuclei to yield qualitative changes. For example the
moment of inertia usually goes only half way from its rigid
body value to its irrotational liquid limit (strong pairing case).
We therefore believe that the physical insights we will develop in
this paper will stay qualitatively correct, even if superfluidity is
included in a later stage. As we mentioned already, this shall be the
subject of future studies.

The paper is organized as follows. In Section 2 the general outline
of the WFM formalism is presented. This formalism is applied to the
model of a harmonic oscillator potential with Q-Q residual interaction
in Section 3: the equations of motion for irreducible tensors are
derived
and analyzed, the energies of collective isoscalar and isovector
excitations are calculated. The method of infinitesimal displacements
is used in Section 4 to find the expressions for the nucleon currents
of the different modes and to display the respective figures.
The magnetic and electric
transition probabilities are calculated in Section 5 with the help of
the linear response theory. Sum rules are analyzed in Section 6. The
most interesting points in the description and understanding of the
scissors mode are discussed in section 7. The scissors mode in
superdeformed nuclei is considered in section 8. The summary of the
main results and concluding remarks are contained in Section 9.
To obtain a general impression of the approach and results one can omit
the sections 4, 5 and 6 in a first reading.

\section{Formulation of the method}

The basis of our method is the Time Dependent Hartree-Fock (TDHF)
equation for the one-body density matrix
$ \rho^{\tau}(\br_1, \br_2, t)=
\langle\br_1|\hat{\rho}^{\tau}(t)|\br_2\rangle $ :
\begin{equation}
\label{TDHF}
 i\hbar \frac{\partial \hat{\rho}^{\tau}}{\partial t}=
\left[ \hat{H}^{\tau} ,\hat{\rho}^{\tau}\right],
\end{equation}
where $ \hat{H}^{\tau} $ is the one-body self-consistent Hamiltonian
depending implicitly on the density matrix and $\tau$ is an isotopic
index.
It is convenient to modify equation (\ref{TDHF}) introducing the
Wigner transform \cite{Wig} of the density matrix
\begin{equation}
\label{f}
 f^{\tau}(\br, \bp, t) = \int d^3s\: \exp(- i\bp
 \cdot \bs/\hbar)\rho^{\tau}(\br+\frac{\bs}{2}, \br-\frac{\bs}{2},t)
\end{equation}
and of the Hamiltonian
\begin{equation}
\label{Hw}
 H_W^{\tau}(\br,\bp)=\int d^3s\: \exp(-i\bp \cdot \bs/\hbar)
(\br+\frac{\bs}{2}\left|\hat{H}^{\tau}\right|\br-\frac{\bs}{2}).
\end{equation}
Using (\ref{f},\ref{Hw}) one arrives \cite{Ring} at

\begin{equation}
\label{fsin}
\frac{\partial f^{\tau}}{\partial t}=
\frac{2}{\hbar}\sin \left(\frac{\hbar}{2}
(\nabla^H_\br \cdot \nabla^f_\bp - \nabla^H_\bp \cdot
\nabla^f_\br)\right) H_W^{\tau} f^{\tau} ,
\end{equation}
 where the upper index on the nabla operator stands for the function
on which this operator acts.
If the Hamiltonian is a sum of a kinetic term and a local
potential $V^{\tau}(\br)$, its Wigner transform is just the classical
version of the same Hamiltonian
\begin{equation}
\label{HwCl}
 H_W^{\tau} = p^2\!/2m + V^{\tau}(\br).
\end{equation}
 Then equation (\ref{fsin}) becomes
\begin{equation}
\label{flocal}
\frac{\partial f^{\tau}}{\partial t}+
\frac{1}{m}{\bf p}\cdot\nabla_{\bf r}f^{\tau}
=\frac{2}{\hbar}\sin\left(\frac{\hbar}{2}\nabla_{\bf r}^V
\cdot\nabla_{\bf p}^f\right)V^{\tau} f^{\tau}.
\end{equation}
 Expanding in powers of $\hbar$ leads to
\begin{equation}
\label{sinexpan}
\frac{\partial f^{\tau}}{\partial t}+
\frac{1}{m}\sum_{i=1}^3 p_i\nabla_i f^{\tau}
-\sum_{i=1}^3 \nabla_iV^{\tau}\nabla_i^pf^{\tau}
+\frac{\hbar^2}{24}\sum_{i,j,k=1}^3 \nabla_i\nabla_j\nabla_kV^{\tau}
\nabla_i^p\nabla_j^p\nabla_k^pf^{\tau}- ... =0.
\end{equation}

 Now we apply the WFM method  to derive a closed
set of dynamical equations for different
multipole moments and other integral
characteristics of the nucleus. This method is
described in detail in Ref. \cite{Bal,BaSc}. Its idea is based on the
virial theorems of Chandrasekhar and Lebovitz \cite{Chand}. It is shown in
\cite{Bal,BaSc}, that by integrating equation (\ref{sinexpan})
over the phase space $\{\bp,\br\}$ with the weights
$x_{i_1}x_{i_2}\ldots x_{i_k}p_{i_{k+1}}\ldots p_{i_{n-1}}p_{i_n}$,
where $k$ runs from $0$ to $n$, one can obtain a closed finite set
of dynamical equations for Cartesian tensors of the rank $n$.
Taking linear combinations of these equations one is able to represent
them through irreducible tensors. However, it is more convenient to
derive the dynamical equations directly for irreducible tensors using
the technique of tensor products \cite{Varshal}. For this it is
necessary to
rewrite the Wigner function equation (\ref{sinexpan}) in terms of
cyclic variables
\begin{eqnarray}
\label{sincyclic}
\frac{\partial f^{\tau}}{\partial t}+
\frac{1}{m}\sum_{\alpha=-1}^1(-1)^{\alpha} p_{-\alpha}\nabla_{\alpha}
f^{\tau}
-\sum_{\alpha=-1}^1(-1)^{\alpha}
\nabla_{-\alpha}V^{\tau}\nabla_{\alpha}^pf^{\tau}
\nonumber\\
+\frac{\hbar^2}{24}\sum_{\alpha,\nu,\sigma=-1}^1(-1)^{\alpha+\nu+\sigma}
 \nabla_{-\alpha}\nabla_{-\nu}\nabla_{-\sigma}V^{\tau}
\nabla_{\alpha}^p\nabla_{\nu}^p\nabla_{\sigma}^pf^{\tau}- ... =0
\end{eqnarray}
 with $$\nabla_{+1}=-\frac{1}{\sqrt 2}(\frac{\partial}{\partial x_1}
+i\frac{\partial}{\partial x_2})~,\quad \nabla_0=\frac{\partial}
{\partial x_3}~,\quad
\nabla_{-1}=\frac{1}{\sqrt 2}(\frac{\partial}{\partial x_1}
-i\frac{\partial}{\partial x_2})~,$$
$$r_{+1}=-\frac{1}{\sqrt 2}(x_1+ix_2)~,\quad r_0=x_3~,\quad
r_{-1}=\frac{1}{\sqrt 2}(x_1-ix_2)$$and the analogous definitions
for $\nabla_{+1}^p~,\quad \nabla_{0}^p~,\quad \nabla_{-1}^p~, $ and
$p_{+1}~,\quad p_{0}~,\quad p_{-1}$.
The required equations shall be obtained by integrating
(\ref{sincyclic})
with different tensor products of $r_{\alpha}$ and $p_{\alpha}$.

\section{Model Hamiltonian, Equations of motion, Eigenfrequencies}

 As outlined in the introduction, the model considered here is a
harmonic oscillator mean field potential with quadrupole-quadrupole
residual interactions. Its microscopic Hamiltonian is
\begin{eqnarray}
\label{Ham}
 H=\sum\limits_{i=1}^A(\frac{\bp_i^2}{2m}+\frac{1}{2}m\omega^2\br_i^2)
+\bar{\kappa}
\sum_{\mu=-2}^{2}(-1)^{\mu}
 \sum\limits_i^Z \sum\limits_j^N
q_{2\mu}(\br_i)q_{2-\mu}(\br_j)
\nonumber\\
+\frac{1}{2}\kappa
\sum_{\mu=-2}^{2}(-1)^{\mu}
\{\sum\limits_{i\neq j}^{Z}
 q_{2\mu}(\br_i)q_{2-\mu}(\br_j)
+\sum\limits_{i\neq j}^{N}
 q_{2\mu}(\br_i)q_{2-\mu}(\br_j)\},
\end{eqnarray}
where the quadrupole operator $q_{2\mu}=\sqrt{16\pi/5}\,r^2Y_{2\mu}$
and $N,Z$ are the numbers of neutrons and protons respectively. The
Hamiltonian is of the standard Bohr-Mottelson type \cite{BM},
however, with an interaction coupling protons and neutrons (constant
$\bar{\kappa}$) and, separately, coupling protons and neutrons among
themselves (constant $\kappa$). This form is appropriate for the
description of the scissors mode.
 The corresponding mean field potentials are
\begin{equation}
\label{poten}
V^{\rm p}(\br,t)=\frac{1}{2}m\,\omega^2r^2+
\sum_{\mu=-2}^{2}(-1)^{\mu}
(\kappa Q_{2\mu}^{\rm p}(t)
+\bar{\kappa}Q_{2\mu}^{\rm n}(t))q_{2-\mu}(\br)
\end{equation}
for protons and
\begin{equation}
\label{noten}
V^{\rm n}(\br,t)=\frac{1}{2}m\,\omega^2r^2+
\sum_{\mu=-2}^2(-1)^{\mu}
(\kappa Q_{2\mu}^{\rm n}(t)
+\bar{\kappa} Q_{2\mu}^{\rm p}(t))q_{2-\mu}(\br)
\end{equation}
for neutrons. The multipole moments $Q_{2\mu}^{\tau}(t)$ are defined as
\begin{equation}
\label{Q2mu}
 Q_{2\mu}^{\tau}(t)=
\int\! d\{\bp,\br\}
q_{2\mu}(\br)f^{\tau}(\br,\bp,t).
\end{equation}
 where
 $\int\! d\{\bp,\br\}\equiv
2 (2\pi\hbar)^{-3}\int\! d^3p\,\int\! d^3r$.
 Introducing the notation
$$
D_{2\mu}^{\rm n}(t)=\kappa Q_{2\mu}^{\rm n}(t)
+\bar{\kappa}Q_{2\mu}^{\rm p}(t), \quad
D_{2\mu}^{\rm p}(t)=\kappa Q_{2\mu}^{\rm p}(t)
+\bar{\kappa}Q_{2\mu}^{\rm n}(t),
$$
one can rewrite the mean fields in a more compact way
\begin{equation}
\label{mfield}
V^{\tau}(\br,t)=\frac{1}{2}m\,\omega^2r^2
+\sum_{\mu=-2}^{2}(-1)^{\mu}
D_{2\mu}^{\tau}(t)q_{2-\mu}(\br).
\end{equation}
 Substituting the spherical functions by the tensor products
$\di r^2Y_{2\mu}=
\sqrt{\frac{3\cdot5}{8\pi}}r_{2\mu}^2~,$
where
$$r^2_{\lambda\mu}\equiv
\{r\otimes r\}_{\lambda\mu}=\sum_{\sigma,\nu}
C_{1\sigma,1\nu}^{\lambda\mu}r_{\sigma}r_{\nu}
$$
and $C_{1\sigma,1\nu}^{\lambda\mu}$ is the Clebsch-Gordan coefficient
(let us recall that the vector $\br$ is a tensor of rank one),
one has
$$V^{\tau}=\frac{1}{2}m\,\omega^2r^2
+\sum_{\mu}(-1)^{\mu}Z_{2\mu}^{\tau}r_{2-\mu}^2.$$
 Here
$$
Z_{2\mu}^{\rm n}=\chi R_{2\mu}^{\rm n}
+\bar{\chi}R_{2\mu}^{\rm p}\,,\quad
Z_{2\mu}^{\rm p}=\chi R_{2\mu}^{\rm p}
+\bar{\chi}R_{2\mu}^{\rm n}\,,$$
$$\chi=6\kappa,\quad\bar\chi=6\bar\kappa,$$
\begin{equation}
\label{Rlmu}
 R_{\lambda\mu}^{\tau}(t)=
\int d\{\bp,\br\}
r_{\lambda\mu}^{2}f^{\tau}(\br,\bp,t).
\end{equation}

 Integration of the equation (\ref{sincyclic}) with the weights
$r_{\lambda\mu}^2~,\,
(rp)_{\lambda\mu}\equiv\{r\otimes p\}_{\lambda\mu}$
and $p_{\lambda\mu}^2$
 yields the following set of equations (it is important to note, that
because (\ref{mfield}) is of
quadratic form, the WFM break off at second order):
\begin{eqnarray}
\label{quadr}
\frac{d}{dt}R_{\lambda\mu}^{\tau}
-\frac{2}{m}L^{\tau}_{\lambda\mu}&=&0,\quad \lambda=0,2
\nonumber\\
\frac{d}{dt}L^{\tau}_{\lambda\mu}
-\frac{1}{m}P_{\lambda\mu}^{\tau}+
m\,\omega^2R^{\tau}_{\lambda \mu}
-2\sqrt5\sum_{j=0}^2\sqrt{2j+1}\{_{2\lambda 1}^{11j}\}
(Z_2^{\tau}R_j^{\tau})_{\lambda \mu}
&=&0,
\quad \lambda=0,1,2
\nonumber\\
\frac{d}{dt}P_{\lambda\mu}^{\tau}
+2m\,\omega^2L^{\tau}_{\lambda \mu}
-4\sqrt5\sum_{j=0}^2\sqrt{2j+1}\{_{2\lambda 1}^{11j}\}
(Z_2^{\tau}L^{\tau}_j)_{\lambda \mu}
&=&0,\quad \lambda=0,2
\end{eqnarray}
 where $\{_{2\lambda 1}^{11j}\}$ is the Wigner
$6j$-symbol.
For the sake of simplicity the time dependence of tensors is not
written out. Further the following
notation is introduced
\begin{equation}
\label{P,L}
 P_{\lambda\mu}^{\tau}(t)=
\int\! d\{\bp,\br\}
p_{\lambda\mu}^{2}f^{\tau}(\br,\bp,t),\quad
 L_{\lambda\mu}^{\tau}(t)=
\int\! d\{\bp,\br\}
(rp)_{\lambda\mu}f^{\tau}(\br,\bp,t).
\end{equation}
It is necessary to say some words about the physical meaning of the
collective variables introduced above. By definition $R_{2\mu}^{\tau}
=Q_{2\mu}^{\tau}/\sqrt6$ and $Q_{2\mu}^{\tau}$ is the quadrupole moment
of the system of particles and $R_{00}^{\tau}=-Q_{00}^{\tau}/\sqrt3$
with $Q_{00}^{\tau}=N^{\tau}<r^2>$ being the mean square radius of the
same system. By analogy with these variables, defined in the coordinate
space, we can say that the variables $P_{2\mu}^{\tau}$ and
$P_{00}^{\tau}$ describe the quadrupole moment and the mean square
radius of the same system in a momentum space. The variables
$L_{\lambda\mu}^{\tau}$ describe the coupling of momentum and
coordinate spaces. To understand their nature it is useful to
recall the definitions \cite{Bal,BaSc} of nuclear density and
mean velocity:
\begin{eqnarray}
 n^{\tau}(\br,t)&=&
\int\! \frac{2d^3p}{ (2\pi\hbar)^3}\,
f^{\tau}(\br,\bp,t),
\nonumber\\
 m n^{\tau}(\br,t) u^{\tau}_i(\br,t)&=&
\int\! \frac{2d^3p}{ (2\pi\hbar)^3}\, p_i f^{\tau}(\br,\bp,t).
\label{densvelo}
\end{eqnarray}
They enter into the definitions (\ref{Rlmu},\ref{P,L})
of irreducible tensors
\begin{eqnarray}
&& R_{\lambda\mu}^{\tau}(t)=
\int\! d^3r\int\! \frac{2d^3p}{ (2\pi\hbar)^3}\,
r_{\lambda\mu}^{2}f^{\tau}(\br,\bp,t)=
\int\! d^3r\,r_{\lambda\mu}^{2}n^{\tau}(\br,t),
\nonumber\\
&& L_{\lambda\mu}^{\tau}(t)=
\int\! d^3r\int\! \frac{2d^3p}{ (2\pi\hbar)^3}\,
(rp)_{\lambda\mu}f^{\tau}(\br,\bp,t)=m
\int\! d^3r\,(ru^{\tau})_{\lambda\mu}n^{\tau}(\br,t).
\label{RLnu}
\end{eqnarray}
The last expression for $L_{\lambda\mu}^{\tau}$ demonstrates in an obvious
way the physical meaning of these variables: being the first order
moments of mean velocities they give information about the
distribution of these velocities in the nucleus. (``First'' means that
velocities are weighted with the coordinate $\br$).
Sometimes, if the motion is comparatively simple, this information
turns out sufficient to completely determine the velocity field
 (see section 4).
In the case of more intricate motions higher order moments are
required for a complete description of velocities \cite{Bal}. In any
case the moments of velocities are a very convenient tool to descibe
the collective motion. For example, the zero order moment of
velocity is nothing more than the linear momentum describing the
nucleus' center of mass motion. One of the first order moments corresponds
to the very well known angular momentum of a nucleus. It is connected with
the variable $L_{1\mu}^{\tau}$ by the following relations:
$$L_{10}^{\tau}=\frac{i}{\sqrt2}I_3^{\tau},\quad L_{1\pm 1}^{\tau}=
\frac{1}{2}(I_2^{\tau}\mp iI_1^{\tau}).$$

  It is convenient to rewrite the equations (\ref{quadr}) in terms
of the isoscalar and isovector variables
$$R_{\lambda\mu}=R_{\lambda\mu}^{\rm n}+R_{\lambda\mu}^{\rm p},\quad
P_{\lambda\mu}=P_{\lambda\mu}^{\rm n}+P_{\lambda\mu}^{\rm p},\quad
L_{\lambda\mu}=L_{\lambda\mu}^{\rm n}+L_{\lambda\mu}^{\rm p},$$
$$\bar R_{\lambda\mu}=R_{\lambda\mu}^{\rm n}-R_{\lambda\mu}^{\rm p}
,\quad
\bar P_{\lambda\mu}=P_{\lambda\mu}^{\rm n}-P_{\lambda\mu}^{\rm p}
,\quad
\bar L_{\lambda\mu}=L_{\lambda\mu}^{\rm n}-L_{\lambda\mu}^{\rm p}.$$
So the equations for the neutron and proton systems are transformed
into isoscalar and isovector ones.
The equations for the isoscalar system are given by
\begin{eqnarray}
\label{isos}
&&\dot R_{00}-2L_{00}/m=0,
\nonumber\\
&&\dot{L}_{00}-P_{00}/m+
m\,\omega^2R_{00}
-2\sqrt{5/3}
[\chi_0 (R_2R_2)_{00}+\chi_1 (\bar R_2\bar R_2)_{00}]
=0,
\nonumber\\
&&\dot P_{00}
+2m\omega^2 L_{00}
-4\sqrt{5/3}
[\chi_0 (R_2 L_2)_{00}+\chi_1 (\bar R_2{\bar L}_2)_{00}]
=0,
\nonumber\\
&&\dot R_{2\mu}-2L_{2\mu}/m=0,
\nonumber\\
&&\dot{L}_{2\mu}-P_{2\mu}/m+
m\,\omega^2R_{2\mu}
-2\sqrt{1/3}
[\chi_0 (R_2R_0)_{2\mu}+\chi_1 (\bar R_2\bar R_0)_{2\mu}]
\nonumber\\
&&\hspace{4.5cm}-\sqrt{7/3}
[\chi_0 (R_2R_2)_{2\mu}+\chi_1 (\bar R_2\bar R_2)_{2\mu}]
=0,
\nonumber\\
&&\dot P_{2\mu}
+2m\omega^2 L_{2\mu}
-4\sqrt{1/3}
[\chi_0(R_2 L_0)_{2\mu}
+\chi_1(\bar R_2{\bar L}_0)_{2\mu}]
\nonumber\\
&&\hspace{4cm}-2\sqrt{7/3}
[\chi_0(R_2 L_2)_{2\mu}+\chi_1 (\bar R_2{\bar L}_2)_{2\mu}]
\nonumber\\
&&\hspace{5cm}+2\sqrt3
[\chi_0(R_2L_1)_{2\mu}+\chi_1 (\bar R_2\bar L_1)_{2\mu}]
=0,
\nonumber\\
&&\dot L_{1\nu}=0.
\end{eqnarray}
 and the ones for the isovector system read:
\begin{eqnarray}
\label{isov}
&&\dot{\bar R}_{00}-2\bar L_{00}/m=0,
\nonumber\\
&&\dot{\bar L}_{00}-\bar P_{00}/m+
m\,\omega^2\bar R_{00}
-2\sqrt{5/3}
\chi(R_2\bar R_2)_{00}
=0,
\nonumber\\
&&\dot{\bar P}_{00}
+2m\omega^2{\bar L}_{00}
-4\sqrt{5/3}
[\chi_0 (R_2{\bar L}_2)_{00}+\chi_1 (\bar R_2 L_2)_{00}]
=0,
\nonumber\\
&&\dot{\bar R}_{2\mu}-2\bar L_{2\mu}/m=0,
\nonumber\\
&&\dot{\bar L}_{2\mu}-\bar P_{2\mu}/m+
m\,\omega^2\bar R_{2\mu}
-2\sqrt{1/3}
[\chi_0 (R_2\bar R_0)_{2\mu}+\chi_1 (\bar R_2R_0)_{2\mu}]
\nonumber\\
&&\hspace{6.5cm}-\sqrt{7/3}
\chi (R_2\bar R_2)_{2\mu}
=0,
\nonumber\\
&&\dot{\bar P}_{2\mu}
+2m\omega^2\bar L_{2\mu}
-4\sqrt{1/3}
[\chi_0(R_2\bar L_0)_{2\mu}
+\chi_1(\bar R_2 L_0)_{2\mu}]
\nonumber\\
&&\hspace{4cm}-2\sqrt{7/3}
[\chi_0(R_2\bar L_2)_{2\mu}+\chi_1 (\bar R_2 L_2)_{2\mu}]
\nonumber\\
&&\hspace{5cm}+2\sqrt3
[\chi_0(R_2\bar L_1)_{2\mu}+\chi_1 (\bar R_2L_1)_{2\mu}]
=0,
\nonumber\\
&&\dot{\bar L}_{1\nu}+\sqrt5\bar{\chi}(R_2\bar R_2)_{1\nu}=0.
\end{eqnarray}
Here $$\chi_0=(\chi+\bar{\chi})/2$$ is an isoscalar strength constant
and $$\chi_1=(\chi-\bar{\chi})/2$$ is the corresponding isovector one.
The last equation of (\ref{isos}) demonstrates the conservation of
the isoscalar angular momentum $L_{1\nu}$. The dynamical
equation for the isovector angular momentum $\bar L_{1\nu}$
(the last equation of (\ref{isov})) describes the relative
(out of phase) motion of the neutron and proton angular
momenta; hence it must be responsible for the scissors mode.

We will need the following algebraic relations:
\begin{eqnarray}
\label{algeb}
(R_2\bar R_2)_{00}&=&\frac{1}{\sqrt5}
(R_{22}\bar R_{2-2}-R_{21}\bar R_{2-1}+R_{20}\bar R_{20}
-R_{2-1}\bar R_{21}+R_{2-2}\bar R_{22}),
\nonumber\\
(R_2\bar R_2)_{20}&=&\sqrt{\frac{2}{7}}
(R_{22}\bar R_{2-2}+R_{21}\bar R_{2-1}/2-R_{20}\bar R_{20}
+R_{2-1}\bar R_{21}/2+R_{2-2}\bar R_{22}),
\nonumber\\
(R_2\bar R_2)_{22}&=&\sqrt{\frac{2}{7}}
(R_{22}\bar R_{20}-\sqrt{\frac{3}{2}}R_{21}\bar R_{21}
+R_{20}\bar R_{22}),
\nonumber\\
(R_2\bar R_2)_{2-2}&=&\sqrt{\frac{2}{7}}
(R_{2-2}\bar R_{20}-\sqrt{\frac{3}{2}}R_{2-1}\bar R_{2-1}
+R_{20}\bar R_{2-2}),
\nonumber\\
(R_2\bar R_2)_{21}&=&\sqrt{\frac{3}{7}}
(R_{22}\bar R_{2-1}
-\frac{1}{\sqrt6}
R_{21}\bar R_{20}
-\frac{1}{\sqrt6}
R_{20}\bar R_{21}+R_{2-1}\bar R_{22}),
\nonumber\\
(R_2\bar R_2)_{2-1}&=&\sqrt{\frac{3}{7}}
(R_{2-2}\bar R_{21}
-\frac{1}{\sqrt6}
R_{2-1}\bar R_{20}
-\frac{1}{\sqrt6}
R_{20}\bar R_{2-1}+R_{21}\bar R_{2-2}),
\nonumber\\
(R_2\bar R_2)_{11}&=&\sqrt{\frac{3}{5}}
(\frac{1}{\sqrt3}
R_{22}\bar R_{2-1}
-\frac{1}{\sqrt2}
R_{21}\bar R_{20}
+\frac{1}{\sqrt2}
R_{20}\bar R_{21}
-\frac{1}{\sqrt3}
R_{2-1}\bar R_{22}),
\nonumber\\
(R_2\bar R_2)_{10}&=&\sqrt{\frac{3}{5}}
(\sqrt{\frac{2}{3}}R_{22}\bar R_{2-2}
-\frac{1}{\sqrt6}
R_{21}\bar R_{2-1}
+\frac{1}{\sqrt6}
R_{2-1}\bar R_{21}
-\sqrt{\frac{2}{3}}R_{2-2}\bar R_{22}),
\nonumber\\
(R_2\bar R_2)_{1-1}&=&\sqrt{\frac{3}{5}}
(\frac{1}{\sqrt3}
R_{21}\bar R_{2-2}
-\frac{1}{\sqrt2}
R_{20}\bar R_{2-1}
+\frac{1}{\sqrt2}
R_{2-1}\bar R_{20}
-\frac{1}{\sqrt3}
R_{2-2}\bar R_{21}),
\nonumber\\
(R_2\bar L_1)_{20}&=&
\frac{1}{\sqrt2}
(R_{21}\bar L_{1-1}-R_{2-1}\bar L_{11}),
\nonumber\\
(R_2\bar L_1)_{22}&=&\sqrt{\frac{2}{3}}
R_{22}\bar L_{10}-
\frac{1}{\sqrt3}
R_{21}\bar L_{11},
\nonumber\\
(R_2\bar L_1)_{2-2}&=&\sqrt{\frac{2}{3}}
R_{2-2}\bar L_{10}-
\frac{1}{\sqrt3}
R_{2-1}\bar L_{1-1},
\nonumber\\
(R_2\bar L_1)_{21}&=&
\frac{1}{\sqrt3}
R_{22}\bar L_{1-1}
+\frac{1}{\sqrt6}
R_{21}\bar L_{10}
-\frac{1}{\sqrt2}
R_{20}\bar L_{11},
\nonumber\\
(R_2\bar L_1)_{2-1}&=&-
\frac{1}{\sqrt3}
R_{2-2}\bar L_{11}-\frac{1}{\sqrt6}
R_{2-1}\bar L_{10}
+\frac{1}{\sqrt2}
R_{20}\bar L_{1-1}.
\end{eqnarray}
With the help of (\ref{algeb}) one can write out in detail the
whole set of 42 coupled equations (including integrals of motion)
for the whole set of isoscalar and isovector variables. There exists
no problem to solve these equations
numerically. However, for the time being we want to simplify the
situation as much as possible what will allow us to get the results
in analytical form and thus will give us a maximum of insight into
the nature of the modes.

1)Let us consider the problem in the small amplitude approximation,
i.e. we only will study small deviations of the system from
equilibrium. Writing all variables as a sum of their
equilibrium value plus a small deviation
$$R_{\lambda\mu}(t)=R_{\lambda\mu}^{eq}+\R_{\lambda\mu}(t),\quad
P_{\lambda\mu}(t)=P_{\lambda\mu}^{eq}+\P_{\lambda\mu}(t),\quad
L_{\lambda\mu}(t)=L_{\lambda\mu}^{eq}+\L_{\lambda\mu}(t),$$
$$\bar R_{\lambda\mu}(t)=\bar R_{\lambda\mu}^{eq}
+\bar \R_{\lambda\mu}(t),\quad
\bar P_{\lambda\mu}(t)=\bar P_{\lambda\mu}^{eq}
+\bar \P_{\lambda\mu}(t),\quad
\bar L_{\lambda\mu}(t)=\bar L_{\lambda\mu}^{eq}
+\bar \L_{\lambda\mu}(t),$$
we linearize the equations of motion in
$\R_{\lambda\mu},\,\P_{\lambda\mu},\,\L_{\lambda\mu}$ and
$\bar \R_{\lambda\mu},\,\bar \P_{\lambda\mu},\,\bar \L_{\lambda\mu}$.

2)We will consider non rotating nuclei, i.e. nuclei with
$L_{1\nu}^{eq}=\bar L_{1\nu}^{eq}=0$.

3)Let us consider axially symmetric nuclei, i.e. nuclei with
$R_{2\pm2}^{eq}=R_{2\pm1}^{eq}=\bar R_{2\pm2}^{eq}=
\bar R_{2\pm1}^{eq}=0$.

4)Finally, we take $\bar R_{20}^{eq}=\bar R_{00}^{eq}=0$. This means
that equilibrium deformation and mean square radius of neutrons are
supposed to be equal to that of protons.

  After all these simplifications the set of equations for the
isoscalar system (\ref{isos}) is transformed into the following set
of linear equations:
\begin{eqnarray}
&&\dot \R_{00}-2\L_{00}/m=0,
\nonumber\\
&&\dot{\L}_{00}-\P_{00}/m+
m\,\omega^2\R_{00}-4\sqrt{1/3}
\chi_0R_{20}^{eq}\,\R_{20}=0,
\nonumber\\
&&\dot \P_{00}
+2m\omega^2 \L_{00}-4\sqrt{1/3}
\chi_0R_{20}^{eq}\,\L_{20}=0,
\nonumber\\
&&\dot \R_{2\pm2}-2\L_{2\pm2}/m=0,
\nonumber\\
&&\dot \R_{2\pm1}-2\L_{2\pm1}/m=0,
\nonumber\\
&&\dot \R_{20}-2\L_{20}/m=0,
\nonumber\\
&&\dot\L_{2\pm2}-\P_{2\pm2}/m+
\left[m\,\omega^2-\sqrt{4/3}\chi_0
(R_{00}^{eq}+\sqrt2R_{20}^{eq})\right]\R_{2\pm2}=0,
\nonumber\\
&&\dot\L_{2\pm1}-\P_{2\pm1}/m+
\left[m\,\omega^2-\sqrt{4/3}\chi_0
(R_{00}^{eq}-R_{20}^{eq}/\sqrt2)\right]\R_{2\pm1}=0,
\nonumber\\
&&\dot\L_{20}-\P_{20}/m+
\left[m\,\omega^2-\sqrt{4/3}\chi_0
(R_{00}^{eq}-\sqrt2R_{20}^{eq})\right]\R_{20}
-\sqrt{4/3}
\chi_0R_{20}^{eq}\,\R_{00}=0,
\nonumber\\
&&\dot \P_{2\pm2}
+2[m\omega^2
-\sqrt{2/3}\chi_0R_{20}^{eq}]\L_{2\pm2}=0,
\nonumber\\
&&\dot\P_{2\pm1}
+2[m\omega^2
+\sqrt{1/6}\chi_0R_{20}^{eq}]\L_{2\pm1}=0,
\nonumber\\
&&\dot \P_{20}
+2[m\omega^2
+\sqrt{2/3}
\chi_0R_{20}^{eq}]\L_{20}
-4\sqrt{1/3}
\chi_0R_{20}^{eq}\,\L_{00}=0,
\nonumber\\
&&\dot \L_{10}=0.
\nonumber\\
&&\dot \L_{1\pm1}=0.
\label{smallsca}
\end{eqnarray}
  The corresponding set of equations for the
isovector system (\ref{isov}) reads
\begin{eqnarray}
&&\dot{\bar\R}_{00}
-2\bar\L_{00}/m=0,
\nonumber\\
&&\dot{\bar\L}_{00}-\bar\P_{00}/m+
m\,\omega^2\bar\R_{00}
-\sqrt{4/3}
\chi R_{20}^{eq}\,\bar\R_{20}
=0,
\nonumber\\
&&\dot{\bar\P}_{00}
+2m\omega^2\bar\L_{00}
-4\sqrt{1/3}
\chi_0R_{20}^{eq}\,\bar\L_{20}
=0,
\nonumber\\
&&\dot{\bar\R}_{2\pm2}-2\bar\L_{2\pm2}/m=0,
\nonumber\\
&&\dot{\bar\R}_{2\pm1}-2\bar\L_{2\pm1}/m=0,
\nonumber\\
&&\dot{\bar\R}_{20}-2\bar\L_{20}/m=0,
\nonumber\\
&&\dot{\bar\L}_{2\pm2}-\bar\P_{2\pm2}/m+
\left[m\,\omega^2
-\sqrt{2/3}\chi R_{20}^{eq}
-\sqrt{4/3}\chi_1R_{00}^{eq}\right]\bar\R_{2\pm2}=0,
\nonumber\\
&&\dot{\bar\L}_{2\pm1}-\bar\P_{2\pm1}/m
+\left[m\,\omega^2
+\sqrt{1/6}\chi R_{20}^{eq}
-\sqrt{4/3}\chi_1R_{00}^{eq}\right]\bar\R_{2\pm1}=0,
\nonumber\\
&&\dot{\bar\L}_{20}-
\bar\P_{20}/m+
\left[m\,\omega^2
+\sqrt{2/3}\chi R_{20}^{eq}
-\sqrt{4/3}\chi_1R_{00}^{eq}\right]\bar\R_{20}
-\sqrt{4/3}
\chi_0\R_{20}^{eq}\,\bar\R_{00}=0,
\nonumber\\
&&\dot{\bar\P}_{2\pm2}
+2[m\omega^2
-\sqrt{2/3}\chi_0R_{20}^{eq}]\bar\L_{2\pm2}=0,
\nonumber\\
&&\dot{\bar\P}_{2\pm1}
+2[m\omega^2
+\sqrt{1/6}\chi_0R_{20}^{eq}]\bar\L_{2\pm1}
\mp\sqrt6\chi_0R_{20}^{eq}\,\bar\L_{1\pm1}=0,
\nonumber\\
&&\dot{\bar\P}_{20}
+2[m\omega^2+\sqrt{2/3}
\chi_0R_{20}^{eq}]\bar\L_{20}
-\sqrt{4/3}
\chi_0R_{20}^{eq}\,\bar\L_{00}=0,
\nonumber\\
&&\dot{\bar\L}_{1\pm1}\pm\sqrt{3/2}\bar{\chi}
R_{20}^{eq}\bar\R_{2\pm1}=0,
\nonumber\\
&&\dot{\bar\L}_{10}=0.
\label{smallvec}
\end{eqnarray}

Due to the approximation 4) the equations for isoscalar and isovector
systems are decoupled. Further, due to the axial symmetry
the angular momentum projection is a good quantum number. As a result,
every set of equations splits into five independent subsets.
For example, the equations (\ref{smallvec}) can be grouped in the
following way:

1) the subset of equations for variables
$\bar\R_{00}, \bar\L_{00}, \bar\P_{00},
\bar\R_{20}, \bar\L_{20}, \bar\P_{20}$ and $\bar\L_{10}$ with
projections $\mu=0$.
The equation for $\bar\L_{10}$ gives the integral of motion. The rest
of equations describes the isovector giant monopole resonance
plus the branch of IVGQR corresponding to $\mu=0 \,(\beta$-mode).

2) the subset of equations for variables
$\bar\R_{22}, \bar\L_{22}, \bar\P_{22}$ with projections $\mu=2$
and the subset of equations for variables
$\bar\R_{2-2}, \bar\L_{2-2}, \bar\P_{2-2}$ with projections $\mu=-2$
describe two degenerate branches of IVGQR corresponding to
$\mu=|2| \,(\gamma$-mode).

3a) the subset of equations for variables
$\bar\R_{21}, \bar\L_{21}, \bar\P_{21}, \bar\L_{11}$ with projections
$\mu=1$
describe the scissors mode plus
the $\mu=1$ branch of IVGQR (transverse shear mode).

3b) the subset of equations for variables
$\bar\R_{2-1}, \bar\L_{2-1}, \bar\P_{2-1}, \bar\L_{1-1}$ with
projections $\mu=-1$
describe the same dynamics as the subset with $\mu=1$,
because the states with $\mu=\pm1$ are degenerate due to the
axial symmetry.

One also should mention that equations (\ref{smallsca},\ref{smallvec})
are equivalent to the RPA equations corresponding to the Hamiltonian
(\ref{Ham}). The RPA equations have partially been solved in
\cite{Suzuki,Bes}.

\subsection{Isoscalar eigenfrequencies}

The dynamics of the isoscalar angular momentum is trivial - no
vibrations, this variable is conserved. However it is necessary to
treat this mode carefully because, being the nonvibrational mode
with zero eigenfrequency, it gives nevertheless a nonzero
contribution to the sum rule (see below). Let us analyze the
isoscalar set of equations with $\mu=\nu=1$ in more detail
\begin{eqnarray}
&&\dot \R_{21}-2\L_{21}/m=0,
\nonumber\\
&&\dot\L_{21}-\P_{21}/m+
\left[m\,\omega^2+\sqrt{4/3}\chi_0
(R_{20}^{eq}/\sqrt2-R_{00}^{eq})\right]\R_{21}=0,
\nonumber\\
&&\dot\P_{21}
+2[m\omega^2
+\sqrt{1/6}\chi_0R_{20}^{eq}]\L_{21}=0,
\nonumber\\
&&\dot \L_{11}=0.
\label{isosca1}
\end{eqnarray}
Imposing the time evolution via $\di{e^{i\Omega t}}$ for all variables
one transforms (\ref{isosca1}) into a set of algebraic equations
with the determinant
\begin{center}
\begin{tabular}{r|c c c c|}
& $i\Omega$ & $-2/m$ & 0 & 0  \\
& $m\,\omega^2+\sqrt{4/3}\chi_0(R_{20}^{eq}/\sqrt{2}-R_{00}^{eq})$ &
$i\Omega$ & $-1/m$ & 0  \\
$\Delta_{is}=$ & 0 & $2m\omega^2+\sqrt{2/3}\chi_0R_{20}^{eq}$ &
$i\Omega$ & $0$ \\
& $0$ & 0 & 0 & $i\Omega$
\end{tabular}
\end{center}
The eigenfrequencies are found from the characteristic equation
$\Delta_{is}=0$ where
\begin{equation}
\label{haracis}
\Delta_{is}=\Omega^2[\Omega^2-4\omega^2-\frac{\chi_0}{m}(
\sqrt6R_{20}^{eq}-\frac{4}{\sqrt3}R_{00}^{eq})].
\end{equation}
Using here the relations $\chi_0=6\kappa_0,\,R_{20}=Q_{20}/\sqrt6$
and $R_{00}=-Q_{00}/\sqrt3$ (where
$Q_{00}=A<r^2>
%=A\frac{3}{5}R^2,\,R=r_0A^{1/3}
$) we rewrite it in
more conventional notation:
\begin{equation}
\label{haracis1}
\Delta_{is}=\Omega^2[\Omega^2-4\omega^2-\frac{6\kappa_0}{m}(
Q_{20}^{eq}+\frac{4}{3}Q_{00}^{eq})].
\end{equation}
For $\kappa_0$ we take the self-consistent value (see Appendix):
$\di{\kappa_0=-\frac{m\bar{\omega}^2}{4Q_{00}}}$,
where $\bar{\omega}^2=\frac{\omega^2}{1+\frac{2}{3}\delta}.$
 Using now the standard
definition of the deformation parameters
$$Q_{20}=Q_{00}\sqrt{5/\pi}\,\beta=
Q_{00}\frac{4}{3}\,\delta$$
we finally obtain
\begin{equation}
\label{haracis2}
\Omega^2[\Omega^2-2\bar{\omega}^2(1+\delta/3)]=0.
\end{equation}
The nontrivial solution of this equation gives the frequency of the
$\mu=1$ branch of the isoscalar GQR
\begin{equation}
\label{omegis}
\Omega^2=\Omega_{is}^2=2\bar{\omega}^2(1+\delta/3).
\end{equation}
In the limit of small deformation this result coincides with that of
\cite{Suzuki}.
The trivial solution $\Omega=\Omega_0=0$ is characteristic of
nonvibrational mode, corresponding to the obvious integral of motion
$\L_{11}=const$ responsible for the rotational degree of freedom.
Another, not so obvious, integral is obtained by a simple
combination of the third and first equations of (\ref{isosca1}):$$
\P_{21}+2[m\omega^2
+\sqrt{1/6}\chi_0R_{20}^{eq}]\frac{m}{2}\R_{21}=const \quad
\longrightarrow \quad
\P_{21}+m^2\bar{\omega}^2
(1+\frac{\delta}{3})\R_{21}=const.$$
Assuming here $\delta=0$ we reproduce our result from ref. \cite{BaSc}
for spherical nuclei, saying that the nuclear density and the Fermi
surface oscillate out of phase.

\subsection{Isovector eigenfrequencies}

The information about the scissors mode is contained in the
set of isovector equations
with $\mu=\nu=1$. Let us analyze it in detail:
\begin{eqnarray}
\label{scis}
&&\dot{\bar\R}_{21}-2\bar\L_{21}/m=0,
\nonumber\\
&&\dot{\bar\L}_{21}-\bar\P_{21}/m
+\left[m\,\omega^2+\sqrt{1/6}\chi R_{20}^{eq}
-\sqrt{4/3}\chi_1R_{00}^{eq}\right]\bar\R_{21}=0,
\nonumber\\
&&\dot{\bar\P}_{21}
+2[m\omega^2+\sqrt{1/6}\chi_0R_{20}^{eq}]\bar\L_{21}
-\sqrt6\chi_0R_{20}^{eq}\,\bar\L_{11}=0,
\nonumber\\
&&\dot{\bar\L}_{11}
+\sqrt{3/2}\bar{\chi}R_{20}^{eq}\bar\R_{21}=0.
\end{eqnarray}
Imposing the time evolution via $\di{e^{i\Omega t}}$ for all variables
one transforms (\ref{scis}) into a set of algebraic equations
with the determinant
\begin{center}
\begin{tabular}{r|c c c c|}
& $i\Omega$ & $-2/m$ & 0 & 0  \\
& $m\,\omega^2+\sqrt{1/6}\chi R_{20}^{eq}-\sqrt{4/3}\chi_1R_{00}^{eq}$ &
$i\Omega$ & $-1/m$ & 0  \\
$\Delta_{iv}=$ & 0 & $2m\omega^2+\sqrt{2/3}\chi_0R_{20}^{eq}$
& $i\Omega$ &
$-\sqrt6\chi_0R_{20}^{eq}$ \\
& $\sqrt{3/2}\bar{\chi}R_{20}^{eq}$ & 0 & 0 & $i\Omega$
\end{tabular}
\end{center}
Again the eigenfrequencies are found from the characteristic equation
$\Delta_{iv}=0$ where
\begin{equation}
\label{harac}
\Delta_{iv}=\Omega^2[\Omega^2-\frac{2}{m}(m\omega^2
+\frac{\chi_0}{\sqrt6}R_{20}^{eq})]
-\frac{2}{m}[
\Omega^2
(m\omega^2
+\frac{\chi}{\sqrt6}R_{20}^{eq}
-\frac{2\chi_1}{\sqrt3}R_{00}^{eq})
-\frac{3}{m}
\bar\chi \chi_0(R_{20}^{eq})^2]
\end{equation}
or in more conventional notation (see the definitions after
eq.(\ref{haracis})):
\begin{equation}
\label{harac1}
\Delta_{iv}=
\Omega^4-\Omega^2[4\omega^2+\frac{8}{m}\kappa_1Q_{00}^{eq}
+\frac{2}{m}(\kappa_1+2\kappa_0)Q_{20}^{eq}]
+\frac{36}{m^2}(\kappa_0-\kappa_1)\kappa_0(Q_{20}^{eq})^2.
\end{equation}
Supposing, as usual, the isovector constant $\kappa_1$
proportional to the isoscalar one, $\kappa_1=\alpha\kappa_0$, and
taking the self-consistent value for $\kappa_0$
we finally obtain
\begin{equation}
\label{harac2}
\Omega^4-2\Omega^2\bar{\omega}^2(2-\alpha)(1+\delta/3)
+4\bar{\omega}^4(1-\alpha)\delta^2=0.
\end{equation}
The solutions of this equation are
\begin{equation}
\label{Omeg2}
\Omega^2_{\pm}=\bar{\omega}^2(2-\alpha)(1+\delta/3)
\pm \sqrt{\bar{\omega}^4(2-\alpha)^2(1+\delta/3)^2
-4\omega^4(1-\alpha)\delta^2}.
\end{equation}
This expression coincides with the result of Hamamoto and Nazarewizh
\cite{Hamam} found in RPA.
The solution $\Omega_+$ gives the frequency $\Omega_{iv}$ of the
$\mu=1$ branch of the isovector GQR.
The solution $\Omega_-$ gives the frequency $\Omega_{sc}$ of the
scissors mode.

 It is worth noticing that in the case $\bar\L_{11}=0$
the set of equations (\ref{scis}) becomes quite similar to
(\ref{isosca1}). Its characteristic equation reduces to the
equation
\begin{equation}
\label{rotzero}
\Omega^3-2\Omega \bar\omega^2(2-\alpha)(1+\delta/3)=0,
\end{equation}
implying that there exists an integral of motion analogous to the
isoscalar one:
$$\bar\P_{21}+m^2\bar\omega^2
(1+\frac{\delta}{3})\bar\R_{21}=const.$$
The nontrivial solution of (\ref{rotzero}) gives the IVGQR frequency for
the case, when rotational degrees of freedom are neglected:
\begin{equation}
\label{IVGQR}
\Omega^2=2\bar\omega^2(2-\alpha)(1+\delta/3).
\end{equation}

Now let us fix the value of the coefficient $\alpha$.
The experimental fact is: the energy of an isovector GQR is
practically two times higher than that of an isoscalar one.
Assuming $\delta=0$ we have
$$\Omega^2_+=\Omega^2_{iv}=2\omega^2(2-\alpha).$$
The simple comparison of this expression with (\ref{omegis}) shows
that the experimental
observation is satisfied by $\alpha=-2$. Then equation
(\ref{Omeg2}) gives the following formulae for both energies:
\begin{eqnarray}
E^2_{iv}=4(1+\delta/3+\sqrt{(1+\delta/3)^2-\frac{3}{4}\delta^2}
\,)(\hbar\bar\omega)^2,
\nonumber\\
E^2_{sc}=4(1+\delta/3-\sqrt{(1+\delta/3)^2-\frac{3}{4}\delta^2}
\,)(\hbar\bar\omega)^2.
\label{Omeg2fin}
\end{eqnarray}

In the limit of small deformations
one can write for IVGQR energy
\begin{equation}
\label{Eniv}
E_{iv}^2\simeq 8(1-\delta/3)(1-\frac{3}{16}\delta^2)(\hbar\omega_0)^2.
\end{equation}
For $\alpha=-2$ formula (\ref{IVGQR}) gives:
$E_{iv}^2\simeq 8(1-\delta/3)(\hbar\omega_0)^2$. Comparing it with
(\ref{Eniv}) one sees
that the influence of rotational degrees of freedom on the IVGQR
energy is very small.

The scissors mode energy in the limit of small deformation is
\begin{equation}
E_{sc}\simeq\sqrt{\frac{3}{2}}\,\delta(1-\delta/2)\hbar\omega_0
\approx\sqrt{\frac{3}{2}}\,\delta\,\hbar\omega_0,
\end{equation}
which is quite close to the result of Hilton \cite{Hilt92}:
$E_{sc}\approx\sqrt{1+0.66}\,\delta\,\hbar\omega_0.$
Taking $\hbar\omega_0=45.2/A^{1/3}$ MeV (what corresponds to
$r_0=1.15$ fm used in \cite{Lipp}), one obtains
$$E_{sc}\approx55.4\,\delta A^{-1/3}\,{\rm MeV},$$
which practically coincides with the result of Lipparini and Stringari
\cite{Lipp}: $E_{sc}\simeq56\,\delta A^{-1/3}$ MeV
obtained with the help of a microscopic approach based on the
evaluation of sum rules.
 Both results are not very far from the experimental \cite{Richter}
value: $E_{sc}\approx 66\delta A^{-1/3}$ MeV.

We now will present the calculation of the flow patterns but the
impatient reader may jump to section 7 for further discussion on
various aspects of the eigenfrequencies.

\section{Flows}

The flow distributions will be calculated with the help of the
method of infinitesimal displacements. This method is based on the
rules of variation of integral quantities of the
object \cite{Chand}. Its detailed description can be
found in \cite{Bal}.

Small variations
$\R^{\tau}_{\lambda\mu}\equiv \delta R^{\tau}_{\lambda\mu}$
and $\L^{\tau}_{\lambda\mu}\equiv \delta L^{\tau}_{\lambda\mu}$
are naturally expressed in terms of variations of $n^{\tau}(\br,t)$
and $u_i^{\tau}(\br,t)$ (see definitions (\ref{densvelo},
\ref{RLnu})):
\begin{eqnarray}
&& \R_{\lambda\mu}^{\tau}(t)=
\int\! d^3r\,
r_{\lambda\mu}^{2}\delta n^{\tau}(\br,t),
\nonumber\\
&& \L_{\lambda\mu}^{\tau}(t)=
m\int\! d^3r\,
[(ru^{\tau}_{eq})_{\lambda\mu}\delta n^{\tau}+
(r\delta u^{\tau})_{\lambda\mu}n^{\tau}_{eq}]=
m\int\! d^3r\,
(r\delta u^{\tau})_{\lambda\mu}n^{\tau}_{eq}.
\label{delRL}
\end{eqnarray}
In the last equation we have supposed that $u^{\tau}_{eq}=0$, i.e.,
there is no motion at equilibrium. The variations
$\delta n$ and $\delta u_i$ are not independent. A relation between
them is obtained by means of the continuity equation \cite{Chand}
$$\delta n=-\sum_{i=1}^3\nabla_i(n\xi_i),\quad
\delta u_i=\frac{\partial \xi_i}{\partial t},$$
where $\xi_i(\br,t)\equiv dx_i$ is an infinitesimal displacement.
Let us represent it in the form of the series:
\begin{equation}
\xi_i(\br,t)=G_i(t)+\sum_{j=1}^3G_{i,j}(t)x_j+\sum_{j,k=1}^3
G_{i,jk}(t)x_jx_k+\sum_{j,k,l=1}^3G_{i,jkl}(t)x_jx_kx_l+\cdots
\label{displ}
\end{equation}
For further use we will conserve only the second term of this
infinite series, neglecting the rest. This procedure is well founded
as explained in \cite{Bal}, so we repeat the most important arguments
very briefly. First of all, it is necessary to notice
that due to the triplanar symmetry of the equilibrium shape of the
nucleus only the tensors $G_{i,j\ldots}$ with an even number of
indices will survive after integration over the volume. Further,
the set of dynamic equations (\ref{quadr}) for the second rank
tensors allows us to describe only rather simple types of motion
with $\xi_i\sim x_i$. To describe a more refined motion with
$\xi_i\sim x_i^3$, one is forced to consider the dynamic equations
for the fourth order moments of the Wigner function (the tensors of
rank four). There is a one-to-one correspondence: the more the motion
is complicated, the larger is the number of moments which must be
considered.

 So we take $\xi_i^{\tau}(\br,t)=\sum_{j=1}^3G_{i,j}^{\tau}(t)x_j$.
It is convenient to introduce the "cyclic" combinations of $\xi_i$
analogously to the cyclic variables in (\ref{sincyclic}):
$$\rho_{+1}^{\tau}=-\frac{1}{\sqrt 2}(\xi_1^{\tau}+i\xi_2^{\tau})~,
\quad \rho_0^{\tau}=\xi_3^{\tau}~,\quad
\rho_{-1}^{\tau}=\frac{1}{\sqrt 2}(\xi_1^{\tau}-i\xi_2^{\tau})$$
and to write them as
$\rho_{\mu}^{\tau}(\br,t)=\sum_{\nu=-1}^{+1}(-1)^{\nu}
S_{\mu,-\nu}^{\tau}(t)r_{\nu}$.
 Then
$$\delta n^{\tau}=-\sum_{i=1}^3\nabla_i(n^{\tau}\xi_i^{\tau})
=-\sum_{\nu=-1}^{+1}(-1)^{\nu}\nabla_{\nu}(n^{\tau}\rho_{-\nu}
^{\tau}),\quad \delta u_{\mu}^{\tau}=\frac{\partial \rho_{\mu}^{\tau}}
{\partial t}=\sum_{\nu=-1}^{+1}(-1)^{\nu}
\dot S_{\mu,-\nu}^{\tau}(t)r_{\nu}.$$
Using these expressions one finds
\begin{eqnarray}
 \R_{\lambda\mu}^{\tau}(t)
&=&-\int d^3r
\sum_{\sigma,\nu}C_{1\sigma,1\nu}^{\lambda\mu}r_{\sigma}r_{\nu}
\sum_{\phi=-1}^{+1}(-1)^{\phi}\nabla_{\phi}(n^{\tau}\rho_{-\phi}
^{\tau})
\nonumber\\
&=&\sum_{\sigma,\nu}C_{1\sigma,1\nu}^{\lambda\mu}
\int d^3r\, n^{\tau}_{eq}(\rho_{\sigma}^{\tau}
r_{\nu}+\rho_{\nu}^{\tau}r_{\sigma})
=2\sum_{\phi,\sigma,\nu}C_{1\sigma,1\nu}^{\lambda\mu}
(-1)^{\phi}\int d^3r\, n^{\tau}_{eq}S_{\sigma,-\phi}^{\tau}
r_{\phi}r_{\nu}
\nonumber\\
&=&2\sum_{k,\kappa}\sum_{\phi,\sigma,\nu}C_{1\sigma,1\nu}^{\lambda\mu}
(-1)^{\phi}S_{\sigma,-\phi}^{\tau}C_{1\phi,1\nu}^{k\kappa}
R_{k\kappa}^{\tau}(eq).
\nonumber
\end{eqnarray}
Now taking into account the axial symmetry ($\kappa=0$) one gets
$$\R_{\lambda\mu}^{\tau}=\frac{2}{\sqrt3}[(\sqrt2 R_{20}^{\tau}
-R_{00}^{\tau})C_{1\mu,10}^{\lambda\mu}S_{\mu,0}^{\tau}-
(\frac{1}{\sqrt2}R_{20}^{\tau}+R_{00}^{\tau})
(C_{1\mu+1,1-1}^{\lambda\mu}S_{\mu+1,-1}^{\tau}+
C_{1\mu-1,11}^{\lambda\mu}S_{\mu-1,1}^{\tau})].$$
Exactly the same derivation for $\L_{\lambda\mu}^{\tau}$ leads to the
following result:
$$\L_{\lambda\mu}^{\tau}=
m\sum_{\sigma,\nu}C_{1\sigma,1\nu}^{\lambda\mu}
\int\! d^3r\, n^{\tau}_{eq}
\dot\rho_{\nu}^{\tau}r_{\sigma}=$$
$$=(-1)^{\lambda}\frac{m}{\sqrt3}[(\sqrt2 R_{20}^{\tau}
-R_{00}^{\tau})C_{1\mu,10}^{\lambda\mu}\dot S_{\mu,0}^{\tau}-
(\frac{1}{\sqrt2}R_{20}^{\tau}+R_{00}^{\tau})
(C_{1\mu+1,1-1}^{\lambda\mu}\dot S_{\mu+1,-1}^{\tau}+
C_{1\mu-1,11}^{\lambda\mu}\dot S_{\mu-1,1}^{\tau})].$$
 We are interested in
$\bar\R_{21}$ and $\bar\L_{11}$.
Remembering that $R_{00}=-Q_{00}/\sqrt3,\,R_{20}=
(\frac{2}{3})^{\frac{3}{2}}Q_{00}\delta$ and $Q_{00}^{\tau}=
\frac{1}{2}Q_{00}$ (due to approximation 4)) we find
$$\bar\R_{2\pm 1}=\frac{1}{3\sqrt2}Q_{00}[
(1-\frac{2}{3}\delta)\bar S_{0,\pm 1}+
(1+\frac{4}{3}\delta)\bar S_{\pm 1,0}], $$
$$\bar\L_{1\pm1}=\frac{m}{6\sqrt2}Q_{00}[
(1-\frac{2}{3}\delta)\dot{\bar S}_{0,\pm 1}-
(1+\frac{4}{3}\delta)\dot{\bar S}_{\pm 1,0}],$$
where
$\bar S_{\sigma,\nu}=S_{\sigma,\nu}^{\rm n}-S_{\sigma,\nu}^{\rm p}$
(and $S_{\sigma,\nu}=S_{\sigma,\nu}^{\rm n}+S_{\sigma,\nu}^{\rm p}$).
Having in mind the $e^{i\Omega t}$ time dependence (vibrational
motion), we can substitute $\dot{\bar S}_{\sigma,\nu}$ by
$i\Omega\bar S_{\sigma,\nu}$. Solving these equations with respect
to $\bar S_{\sigma,\nu}$, we have
$$\bar S_{0,1}=\frac{3}{\sqrt2}[\bar\R_{21}-\frac{2i}{m\Omega}
\bar\L_{11}]/[Q_{00}(1-\frac{2}{3}\delta)],\quad
\bar S_{1,0}=\frac{3}{\sqrt2}[\bar\R_{21}+\frac{2i}{m\Omega}
\bar\L_{11}]/[Q_{00}(1+\frac{4}{3}\delta)].$$
Now we use the set of equations (\ref{scis}) to find that
$\bar\L_{11}=-\frac{i}{\Omega}m\bar\omega^2\delta(1-\alpha)
\bar\R_{21}$
and, as a result,
$$\bar\R_{21}\mp\frac{2i}{m\Omega}\bar\L_{11}=
[1\mp 2\frac{\bar\omega^2}{\Omega^2}(1-\alpha)\delta]\bar\R_{21}.$$
Introducing the notation
\begin{eqnarray}
A=\frac{3}{\sqrt2}[1-2\frac{\bar\omega^2}{\Omega^2}(1-\alpha)\delta]
/[Q_{00}(1-\frac{2}{3}\delta)],
\nonumber\\
B=\frac{3}{\sqrt2}[1+2\frac{\bar\omega^2}{\Omega^2}(1-\alpha)\delta]
/[Q_{00}(1+\frac{4}{3}\delta)],
\label{AiB}
\end{eqnarray}
we finally get
$$\bar S_{0,1}=A\bar\R_{21},\quad\bar S_{1,0}=B\bar\R_{21}.$$
A similar analysis of $\bar\R_{2-1}$ and $\bar\L_{1-1}$
allows us to write immediately
$$\bar S_{0,-1}=A\bar\R_{2-1},\quad\bar S_{-1,0}=B\bar\R_{2-1}.$$
So we have for the "cyclic" displacements:
$$\bar\rho_{+1}=\bar S_{1,0}r_0=B\bar\R_{21}x_3,\quad
\bar\rho_{-1}=\bar S_{-1,0}r_0=B\bar\R_{2-1}x_3,$$
$$\bar\rho_{0}=-\bar S_{0,1}r_{-1}-\bar S_{0,-1}r_{+1}
=\sqrt2 A(\bar\J_{13}x_1+\bar\J_{23}x_2),$$
where
$\bar\J_{13}=(\bar\R_{2-1}-\bar\R_{21})/2,\quad
\bar\J_{23}=i(\bar\R_{2-1}+\bar\R_{21})/2.$
By definition, the variable $\J_{ij}^{\tau}$ is a small variation
of the tensor $J_{ij}^{\tau}=\int\! d\{\bp,\br\}
x_ix_jf^{\tau}(\br,\bp,t).$
Cartesian displacements are found by elementary means:
$$\bar\xi_1=\frac{1}{\sqrt2}(\bar\rho_{-1}-\bar\rho_{+1})=\sqrt2 B
\bar\J_{13}x_3,\quad
\bar\xi_2=\frac{i}{\sqrt2}(\bar\rho_{-1}+\bar\rho_{+1})=\sqrt2 B
\bar\J_{23}x_3,$$
$$\bar\xi_3=\bar\rho_{0}=
\sqrt2 A(\bar\J_{13}x_1+\bar\J_{23}x_2).$$

 Let us analyze the picture of displacements in the plane $x_1=0$
(the picture in the plane $x_2=0$ must be exactly the same due to
axial symmetry). Knowing the infinitesimal displacements
\begin{equation}
\label{displac}
\bar\xi_2\equiv dy=\sqrt2 B\bar\J_{23}x_3,\quad
\bar\xi_3\equiv dz=\sqrt2 A\bar\J_{23}x_2,
\end{equation}
we can derive the
differential equation for the flow
$$\frac{dy}{dz}=\frac{B}{A}\frac{z}{y}\quad\longrightarrow \quad
ydy-\frac{B}{A}zdz=0.$$
Integrating this equation we find
$$y^2+\beta z^2= const\equiv c\quad \longrightarrow \quad
\frac{y^2}{c}+\frac{z^2}{c/\beta}=1,$$
where $\beta=-B/A.$ Depending on the sign of $\beta$ this curve will
be either an ellipse or a hyperbola. So, it is necessary to study
carefully the structure of the coefficient $\beta$. It is
convenient to study the coefficients $A$ and $B$ separately.

 Let us investigate at first the case of the {\bf scissors mode}.
Taking $\Omega=\Omega_{sc}$ and $\alpha=-2$ we have
\begin{eqnarray}
A=\frac{3(1+\delta/3-\sqrt{(1+\delta/3)^2-\frac{3}{4}\delta^2}
-\frac{3}{2}\delta)}
{\sqrt2Q_{00}(1-\frac{2}{3}\delta)
(1+\delta/3-\sqrt{(1+\delta/3)^2-\frac{3}{4}\delta^2})}.
\label{flowA}
\end{eqnarray}
The bounds of possible values of $\delta$ are
determined by the natural requirements $\omega_{x,y,z}^2>0$. They give
(see Appendix):
\begin{equation}
-\frac{3}{4}\,<\,\delta\,<\,\frac{3}{2}.
\label{bounsc}
\end{equation}
It is easy to check that inside of these bounds the square root
$\sqrt{(1+\delta/3)^2-\frac{3}{4}\delta^2}$ is real and the
denominator of
expression (\ref{flowA}) is always positive. The sign of the numerator
depends on the sign of $\delta$. Elementary analysis of
the function $$F(\delta)=1+\delta/3
-\sqrt{(1+\delta/3)^2-\frac{3}{4}\delta^2}-\frac{3}{2}\delta$$
shows that $F(\delta)=0$ at $\delta=0$ and its derivative
$F^{'}(\delta)$ is negative in this point. This means that
$F(\delta)>0$ for $\delta<0$ and $F(\delta)<0$ for $\delta>0.$
 Hence, $A>0$ for $\delta<0$ and $A<0$ for $\delta>0.$
Analogous analysis of $\delta$-dependence of $B$ shows that
 $B<0$ for $\delta<0$
and $B>0$ for $\delta>0.$ So, we can conclude that $\beta>0$ for any
$\delta$ and the currents in the case of the scissors mode are described
by ellipses.

\begin{figure}
\begin{center}
\epsfig{file=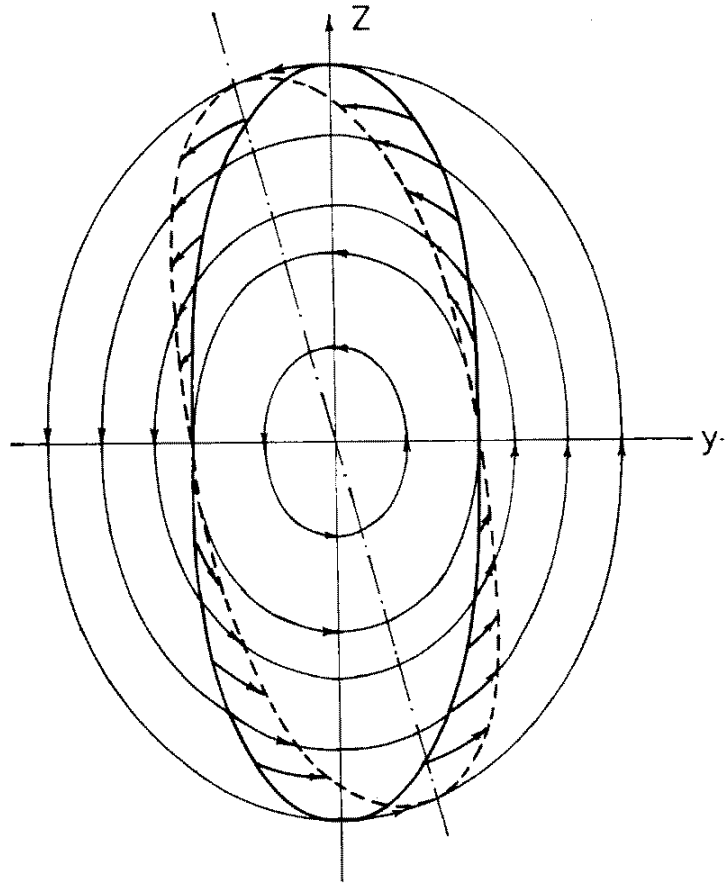,width=10cm}
\end{center}
\caption{Schematic picture of isovector displacements for the
scissors mode. Thin ellipses are the lines of currents. The thick
oval is the initial position of the nucleus' surface (common for
protons and neutrons). The dashed
oval is the final position of the protons' (or neutrons') surface as
a result of infinitesimal displacements shown by the arrows.
\label{fig1}}
\end{figure}

 Let us consider the limit of small $\delta$ for an illustration. We
have
\begin{eqnarray}
\beta&=&-\frac{1-\frac{2}{3}\delta}{1+\frac{4}{3}\delta}\,\frac
{1+\delta/3-\sqrt{(1+\delta/3)^2-\frac{3}{4}\delta^2}+
\frac{3}{2}\delta}
{1+\delta/3-\sqrt{(1+\delta/3)^2-\frac{3}{4}\delta^2}-
\frac{3}{2}\delta}
\nonumber\\
&\simeq&(1-2\delta)\,
\frac{1+\delta/3+\frac{1}{4}\delta}{1+\delta/3-\frac{1}{4}\delta}
\approx(1-2\delta)(1+\frac{1}{2}\delta)
\approx(1-\frac{3}{2}\delta).
\nonumber
\end{eqnarray}
So, for $\delta>0$ the short semiaxis of the current ellipse is
$\Y^2=c$ and the long semiaxis is $\Z^2=c/\beta\simeq
c(1+\frac{3}{2}\delta)$. The eccentricity of this ellipse is
$$e_{cur}^2=1-\frac{\Y^2}{\Z^2}=1-\beta=\frac{3}{2}\delta,$$
which must be compared with the eccentricity of ellipsoid
corresponding to the shape of the mean field:
$$e_{body}^2=1-\frac{<y^2>}{<z^2>}=
\frac{2\delta}{1+\frac{4}{3}\delta}\simeq 2\delta.$$
Thus, the field of currents and the shape of a nucleus are described
by prolate ellipsoides, their long (and short) semiaxes being
disposed along the same coordinate axis. Figure~\ref{fig1} illustrates
the situation schematically. The displacements and the difference in
eccentricities are exaggerated on purpose to demonstrate clearly the
essential features of the motion corresponding to the scissors mode.
One can easily see that its main constituent is a rotation (out of
phase rotation of neutrons and protons). It is also
seen that the rotation is accompanied by the distortion of the
nucler shape - at least it is evident that the long semiaxis
becomes smaller.

 Hence, the real motion of the scissors mode is a mixture of
rotational and irrotational ones.
To get a quantitative measure for the contribution of each kind of
motion, it is sufficient to write the displacement $\vec\xi$ as the
respective superposition \cite{Zaw}:
$$\vec\xi=a\vec e_x\times\vec r+b\nabla(yz)=a(0,-z,y)+b(0,z,y).$$
Comparing the components $\xi_y=(b-a)z,\,\, \xi_z=(b+a)y$ with $\xi_2,
\,\,\xi_3$ in (\ref{displac}) we immediately find
$$b-a=\sqrt2\bar J_{23}B,\quad b+a=\sqrt2\bar J_{23}A \quad
\longrightarrow \quad a=D(1+\beta),\quad b=D(1-\beta),$$
where $D=\bar J_{23}A/\sqrt2.$ So in the considered example we have
$$a=2D(1-\frac{3}{4}\delta),\quad b=\frac{3}{2}D\delta,\quad
b/a\simeq\frac{3}{4}\delta(1+\frac{3}{4}\delta)\approx
\frac{3}{4}\delta.$$
This value of the ratio $b/a$ has the same order
of magnitude as another measure of an
"admixture": the ratio $B(M1)_{iv}/B(M1)_{sc}\simeq\frac{3\sqrt3}{4}
\delta$ (see the subsections 7.2 and 7.4). We also have to mention
the following interesting fact. As we know (see section 7.3), in
the absence of the Fermi Surface Deformation (FSD) the scissors mode
is a zero frequency mode. Calculating (with the help of formulae
(\ref{AiB})) the ratio $B/A$ for $\Omega=0$ we find that ${\di \beta
=\frac{1-\frac{2}{3}\delta}{1+\frac{4}{3}\delta}\simeq 1-2\delta}$.
Hence, the eccentricity of the current ellipse ($e_{cur}^2=2\delta$)
coincides with that of the body ellipsoid. As a result, the lines of
flows are tangential to the nuclear surface, i.e. the motion goes
without any separation of neutron and proton surfaces (in agreement
with the results of papers \cite{Lipp9,ZawSp}).
 Looking at Fig. \ref{fig1}
we can conclude that the inclusion of FSD inevitably leads to the
separation of neutrons and protons, what justifies the name of
scissors mode, independently of how large the separation is.

\begin{figure}
\begin{center}
\epsfig{file=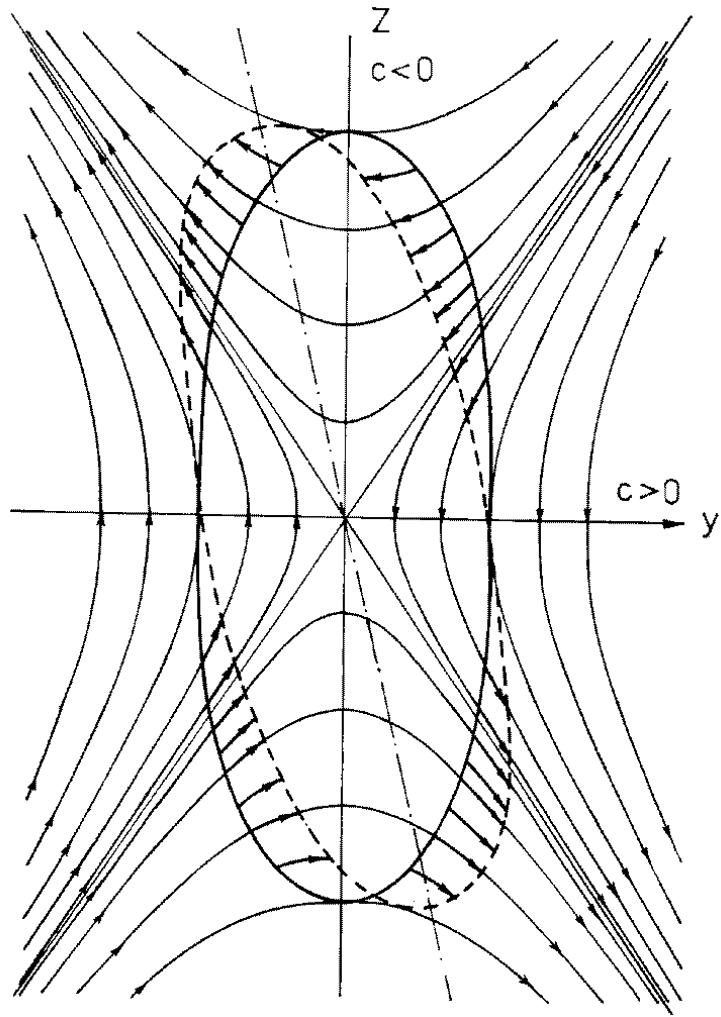,width=10cm}
\end{center}
\caption{Schematic picture of isovector displacements for the
high-lying mode (IVGQR). The lines of currents are shown by thin
curves (hyperbolae). The thick oval is the initial position of the
nucleus' surface (common for protons and neutrons). The dashed
oval is the final position of the protons'
(or neutrons') surface as a result of infinitesimal displacements
shown by the arrows.\label{fig2}}
\end{figure}

Let us investigate now the structure of flows for the {\bf high-lying
mode (IVGQR)}. Taking in (\ref{AiB}) $\Omega=\Omega_{iv}$ and
$\alpha=-2$ we find
$$
A=\frac{3(1+\delta/3+\sqrt{(1+\delta/3)^2-\frac{3}{4}\delta^2}
-\frac{3}{2}\delta)}
{\sqrt2Q_{00}(1-\frac{2}{3}\delta)
(1+\delta/3+\sqrt{(1+\delta/3)^2-\frac{3}{4}\delta^2})}.
$$
It is obvious that the denominator is positive in the above-mentioned
bounds: $-\frac{3}{4}<\delta<\frac{3}{2}$. Elementary
calculations show that the numerator is equal to zero at $\delta=3/2$,
being positive for $\delta<3/2$ and negative for $\delta>3/2$. Hence,
$A>0$ for $\delta<3/2$ and $A<0$ for $\delta>3/2$. An analogous
analysis of the expression
$$
B=\frac{3(1+\delta/3+\sqrt{(1+\delta/3)^2-\frac{3}{4}\delta^2}
+\frac{3}{2}\delta)}
{\sqrt2Q_{00}(1+\frac{4}{3}\delta)
(1+\delta/3+\sqrt{(1+\delta/3)^2-\frac{3}{4}\delta^2})}.
$$
shows that $B$ is equal to zero at $\delta=-3/4$, being positive for
$\delta>-3/4$ and negative for $\delta<-3/4$. So we can conclude that
$\beta<0$ for $-3/4<\delta<3/2$. Hence, the currents in the case of
IVGQR are described by a hyperbola.
 As usual, the situation is illustrated for the
case with small $\delta$. We have
$$\beta=-\frac{1-\frac{2}{3}\delta}{1+\frac{4}{3}\delta}\,
\frac{1+\delta/3+\sqrt{(1+\delta/3)^2+
\frac{3}{4}\delta^2}+\frac{3}{2}\delta}
{1+\delta/3+\sqrt{(1+\delta/3)^2-
\frac{3}{4}\delta^2}-\frac{3}{2}\delta}
\simeq
-(1-2\delta)(1+\frac{3}{2}\delta)
\approx-(1-\frac{1}{2}\delta).$$
The family of curves $y^2-(1-\frac{1}{2}\delta)z^2=c$ is displayed
schematically in Fig. \ref{fig2}. The most remarkable property of the
current lines is seen with one glance: they are nonclosed,
demonstrating the typical sample (see, for example, first page of
\cite{Greiner}) of irrotational motion. Nevertheless, the
final position of the surface of the proton (or neutron) system
looks, as if this system was
rotated on the whole, the length of the big semiaxis being increased.
That is, from the outside one again sees practically the same picture,
as in the case of the scissors mode: rotation plus distortion. This
is a curious property of the shear motion and it justifies the second
name of the $K^{\pi}=1^+$ branch of IVGQR as "the high energy scissors
mode" \cite{Hilt92,Lo2000}. The quantitative
contributions for the two kinds of motion to the IVGQR are
$$a=\frac{1}{2}D\delta,\quad b=2D(1-\frac{1}{4}\delta),\quad
a/b\simeq\frac{1}{4}\delta(1+\frac{1}{4}\delta)\approx
\frac{1}{4}\delta.$$

Concluding the comparison of the scissors mode current with that of
IVGQR it is worth noticing that two principally different
types of infinitesimal displacements result approximately in the same
change of the nuclear surface position.

\section{Linear response and transition probabilities}

A direct way of calculating the reduced transition probabilities
is provided by the theory of linear response of a system to a weak
external field
\begin{equation}
\label{extf}
\hat O(t)=\hat O\,exp(-i\Omega t)+\hat O^{\dagger}\,exp(i\Omega t).
\end{equation}
A convenient form of the response theory is e.g. given by Lane
\cite{Lan}.
The matrix elements of the operator $\hat O$ obey the relationship
\begin{equation}
\label{matel}
|<\psi_a|\hat O|\psi_0>|^2=
\hbar\lim_{\Omega\to\Omega_a}(\Omega-\Omega_a)
\overline{<\psi'|\hat O|\psi'>\exp(-i\Omega t)},
\end{equation}
where $\psi_0$ and $\psi_a$ are the stationary wave functions of
unperturbed ground and excited states; $\psi'$ is the wavefunction
of the perturbed ground state, $\Omega_a=(E_a-E_0)/\hbar$ are the
normal frequencies, the bar means averaging over a time interval much
larger than $1/\Omega$, $\Omega$ being the frequency of the external
field $\hat O(t)$.

 To use formula (\ref{matel}) in the frame of our method, one must
solve two problems \cite{Bal}:

(1) to express the matrix element $<\psi'|\hat O|\psi'>$ in terms of
collective variables of the system,

(2) to find the solution of the dynamic equations for these variables
in the presence of the external field.

The first problem is solved with the help of the formula for the Wigner
transformation of a product of two operators \cite{Ring}
\begin{eqnarray}
\label{matelW}
<\psi'|\hat O|\psi'>&=&\int d^3r \int d^3r'\rho(\br,\br',t)
\hat O(\br',\br)
\\
&=&\int d^3r \int \frac{4d^3p}{ (2\pi\hbar)^3}\,
\exp \left(\frac{\hbar}{2i}
(\nabla^O_\br \cdot \nabla^f_\bp - \nabla^O_\bp \cdot
\nabla^f_\br)\right) O_W(\br,\bp) f(\br,\bp,t).
\nonumber
\end{eqnarray}

To deal with the second problem we add the field (\ref{extf}) to the
mean field potential (\ref{mfield}). The equation for the Wigner
function (\ref{fsin}) is then modified by the term
\begin{equation}
\label{fsinext}
F_{ext}=\frac{2}{\hbar}\sin \left(\frac{\hbar}{2}
(\nabla^O_\br \cdot \nabla^f_\bp - \nabla^O_\bp \cdot
\nabla^f_\br)\right) \left(O_W\exp(-i\Omega t)
+O_W^*\exp(i\Omega t)\right) f.
\end{equation}
Proceeding in the same way as before one obtains equations for all
collective variables needed to calculate
$<\psi'|\hat O|\psi'>$. The only new element now is the presence of
the term $F_{ext}$ that makes the equations for the moments
inhomogeneous.

\subsection{B(M1)-factors}

To calculate the magnetic transition probability, it is necessary
to excite the system with the following external field:
$$\hat O_{\lambda\mu'}=-i\frac{e\hbar}{mc}\frac{1}{\lambda+1}\nabla
(r^{\lambda}Y_{\lambda\mu'})\cdot[\br\times\nabla].$$
We are interested in the dipole operator ($\lambda=1$). In the cyclic
coordinates it looks like
\begin{equation}
\label{Magnet1}
\hat O_{1\mu'}=-\frac{e\hbar}{2mc}\sqrt{\frac{3}{2\pi}}\sum_{\nu,\sigma}
C_{1\nu,1\sigma}^{1\mu'}r_{\nu}\nabla_{\sigma},\quad
\hat O_{1\mu'}^{\dagger}=-\hat O_{1\mu'}^*=(-1)^{\mu'}\hat O_{1-\mu'}.
\end{equation}
Its Wigner transformation is
$$(O_{1\mu'})_W=\gamma\sum_{\nu,\sigma}
C_{1\nu,1\sigma}^{1\mu'}r_{\nu}p_{\sigma}=\gamma(rp)_{1\mu'},$$
where
$\di{\gamma=-i\frac{e}{2mc}\sqrt{\frac{3}{2\pi}}}$.
 For its matrix element we have
\begin{equation}
\label{psiO}
<\psi'|\hat O_{1\mu'}|\psi'>=\gamma L_{1\mu'}^{\rm p}=
\frac{\gamma}{2}(L_{1\mu'}-\bar L_{1\mu'})=
\frac{\gamma}{2}(\L_{1\mu'}-\bar \L_{1\mu'}).
\end{equation}
 Here we have taken into account that
$L_{1\mu'}^{eq}=\bar L_{1\mu'}^{eq}=0$.
The contribution of $\hat O_{1\mu'}(t)$ to the equation for the Wigner
function is
$$F_{ext}=\gamma\left(F_{\mu'}\exp^{-i\Omega t}+
(-1)^{\mu'}F_{-\mu'}\exp^{i\Omega t}\right)$$
with
$$F_{\mu'}=\sum_{\nu\sigma}C_{1\nu,1\sigma}^{1\mu'}
[p_{\sigma}\nabla_{\nu}^p-r_{\nu}\nabla_{\sigma}^r]f^{\rm p}.$$
Integration of $F_{\mu'}$ with the weights
$r_{\lambda\mu}^2~,\, (rp)_{\lambda\mu}$ and $p_{\lambda\mu}^2$
 yields
$$\int d\{\bp,\br\}r_{\lambda\mu}^2F_{\mu'}=
2\sqrt{3(2\lambda+1)}\sum_{k,\pi}C_{\lambda\mu,1\mu'}^{k\pi}
\{^{11\lambda}_{k11}\}R_{k\pi}^{\rm p}(eq),$$
$$\int d\{\bp,\br\}(rp)_{\lambda\mu}F_{\mu'}=
\sqrt{3(2\lambda+1)}\sum_{k,\pi}
[(-1)^{\lambda}+(-1)^k]C_{\lambda\mu,1\mu'}^{k\pi}
\{^{11\lambda}_{k11}\}L_{k\pi}^{\rm p}(eq),$$
$$\int d\{\bp,\br\}p_{\lambda\mu}^2F_{\mu'}=
2\sqrt{3(2\lambda+1)}\sum_{k,\pi}C_{\lambda\mu,1\mu'}^{k\pi}
\{^{11\lambda}_{k11}\}P_{k\pi}^{\rm p}(eq).$$
  A simple analysis of these expressions shows that the
external field modifies only the proton part of the set of equations
(\ref{quadr}) with $\lambda=2$:
\begin{eqnarray}
\label{P2}
\frac{d}{dt}R_{2\mu}^{\rm p}-\cdots =
-\gamma\sqrt3\left[C_{2\mu,1\mu'}^{2\mu+\mu'}R_{2\mu+\mu'}^{\rm p}(eq)
\exp^{-i\Omega t}
+(-1)^{\mu'}C_{2\mu,1-\mu'}^{2\mu-\mu'}R_{2\mu-\mu'}^{\rm p}(eq)
\exp^{i\Omega t}\right],
\nonumber\\
\frac{d}{dt}L^{\rm p}_{2\mu}-\cdots=0,\hspace{11.8cm}\,
\nonumber\\
\frac{d}{dt}P_{2\mu}^{\rm p}+\cdots=
-\gamma\sqrt3\left[C_{2\mu,1\mu'}^{2\mu+\mu'}P_{2\mu+\mu'}^{\rm p}(eq)
\exp^{-i\Omega t}
+(-1)^{\mu'}C_{2\mu,1-\mu'}^{2\mu-\mu'}P_{2\mu-\mu'}^{\rm p}(eq)
\exp^{i\Omega t}\right].\,\,
\end{eqnarray}
We put here $L_{2\mu}^{\rm p}(eq)=0$.
The modifications of the respective isoscalar and isovector
equations are obvious.

The $\mu'=0$ component of the external field does not disturb a
nucleus due to its axial symmetry. Let us consider the case of
$\mu'=1$. The set of equations (\ref{P2}) reads
\begin{eqnarray}
\label{inhomiv}
&&\dot{R}_{2\mu}^{\rm p}-\cdots =
\gamma\sqrt{3/8}R_{20}^{eq}\left[\delta_{\mu,-1}
\exp^{-i\Omega t}
+\delta_{\mu,1}\exp^{i\Omega t}\right],
\nonumber\\
&&\dot{L}_{2\mu}^{\rm p}-\cdots=0,
\nonumber\\
&&\dot{P}_{2\mu}^{\rm p}+\cdots=
\gamma\sqrt{3/8}P_{20}^{eq}\left[\delta_{\mu,-1}
\exp^{-i\Omega t}
+\delta_{\mu,1}\exp^{i\Omega t}\right].
\end{eqnarray}
We have used herein the relations $R_{20}^{\rm p}(eq)=R_{20}^{eq}/2,\,
P_{20}^{\rm p}(eq)=P_{20}^{eq}/2$ which hold true due to
approximation 4).

Now, in accord with formula (\ref{psiO}), we have to find
the tensors $\bar \L_{11}$ and $\L_{11}$.
The tensor $\bar \L_{11}$ is found by solving
the modified (as in (\ref{inhomiv})) set of
equations (\ref{scis}):
\begin{eqnarray}
\label{scisinh}
\dot{\bar\R}_{21}-2\bar\L_{21}/m&=&
-\gamma\sqrt{3/8}R_{20}^{eq}\exp^{i\Omega t},
\nonumber\\
\dot{\bar\L}_{21}-\bar\P_{21}/m
+\left[m\,\omega^2+\sqrt{1/6}\chi R_{20}^{eq}
-\sqrt{4/3}\chi_1R_{00}^{eq}\right]\bar\R_{21}&=&0,
\nonumber\\
\dot{\bar\P}_{21}
+2[m\omega^2+\sqrt{1/6}\chi_0R_{20}^{eq}]\bar\L_{21}
-\sqrt6\chi_0R_{20}^{eq}\,\bar\L_{11}&=&
-\gamma\sqrt{3/8}P_{20}^{eq}\exp^{i\Omega t},
\nonumber\\
\dot{\bar\L}_{11}
+\sqrt{3/2}\bar{\chi}R_{20}^{eq}\bar\R_{21}&=&0.
\end{eqnarray}
It is clear that the time dependence of all variables
must be $\exp^{i\Omega t}$. The required variable is determined
by the ratio of two determinants
$$\bar\L_{11}=\frac{\Delta_{\bar\L}}{\Delta_{iv}}\exp^{i\Omega t},$$
where $\Delta_{iv}$ is defined by (\ref{harac1}) and
$$\Delta_{\bar\L}=\frac{3}{4}\gamma\bar\chi R_{20}^{eq}
\left[R_{20}^{eq}(2\omega^2+\sqrt{2/3}\frac{\chi_0}{m}R_{20}^{eq}
-\Omega^2)+\frac{2}{m^2}P_{20}^{eq}\right].$$
At equilibrium the set of dynamic equations
(\ref{quadr}) considerably simplify
turning into the set of equations of equilibrium.
Taking into account one of them
$$\frac{1}{m}P_{20}^{eq}=m\omega^2R_{20}^{eq}
-\frac{2}{\sqrt3}\chi_0R_{20}^{eq}R_{00}^{eq}
+\frac{2}{\sqrt6}\chi_0(R_{20}^{eq})^2$$
we obtain
$$\Delta_{\bar\L}=\frac{3}{4}\gamma\bar\kappa Q_{20}^2
[4\omega^2+\frac{\kappa_0}{m}(6Q_{20}+8Q_{00})-\Omega^2].$$

Looking at the isoscalar counterpart of the set of equations
(\ref{scisinh})
\begin{eqnarray}
\dot \R_{21}-2\L_{21}/m&=&
\gamma\sqrt{3/8}R_{20}^{eq}\exp^{i\Omega t},
\nonumber\\
\dot\L_{21}-\P_{21}/m+
\left[m\,\omega^2+\sqrt{4/3}\chi_0
(R_{20}^{eq}/\sqrt2-R_{00}^{eq})\right]\R_{21}&=&0,
\nonumber\\
\dot\P_{21}
+2[m\omega^2
+\sqrt{1/6}\chi_0R_{20}^{eq}]\L_{21}&=&
\gamma\sqrt{3/8}P_{20}^{eq}\exp^{i\Omega t},
\nonumber\\
\dot \L_{11}&=&0
\label{isoscainh}
\end{eqnarray}
one easily finds that the isoscalar tensor $\L_{11}=0$.

Writing now the determinant $\Delta_{iv}$ as
\begin{equation}
\label{Delta}
\Delta_{iv}=(\Omega^2-\Omega_{iv}^2)(\Omega^2-\Omega_{sc}^2),
\end{equation}
we easily can find the limit (\ref{matel}).
For the case where $|\psi_a>=|\psi_{sc}>$ we have
$$|<sc|\hat O_{11}|0>|^2=
-\frac{\gamma}{2}\hbar\Delta_{\bar\L}(\Omega_{sc})/
[(\Omega^2_{sc}-\Omega_{iv}^2)2\Omega_{sc}].$$
   Applying the standard values of parameters
$$\kappa_1=\alpha\kappa_0,\quad 4\kappa_0Q_{00}=-m\bar\omega^2,
\quad \kappa_0Q_{20}=-\frac{\delta}{3}m\bar\omega^2$$
we arrive at a rather
complicated function of the deformation parameter $\delta$
\begin{eqnarray}
\label{scimat}
|<sc|\hat O_{11}|0>|^2=\frac{1-\alpha}{8\pi}m\bar\omega^2
Q_{00}\delta^2[E_{sc}^2-2(1+\delta/3)(\hbar\bar\omega)^2]/
[E_{sc}(E^2_{sc}-E_{iv}^2)]\,\mu_N^2,\quad
\end{eqnarray}
where $\di{\mu_N=\frac{e\hbar}{2mc}}$ and
$E_{sc}^2$ and $E_{iv}^2$ are given by (\ref{Omeg2}).
For small values of $\delta$
this expression is considerably simplified. Assuming $\alpha=-2$ one
finds the formula
$$
|<sc|\hat O_{11}|0>|^2/\mu_N^2\simeq
\sqrt{\frac{3}{2}}
\frac{Q_{00}^0}{16\pi}\frac{m\omega_0}{\hbar}
\frac{\delta}{1+\delta/6}
\approx
\sqrt{\frac{3}{2}}
\frac{Q_{00}^0}{16\pi}\frac{m\omega_0}{\hbar}\,\delta
,$$
demonstrating the familiar \cite{Zaw} linear dependence on $\delta$.

For the case $|\psi_a>=|\psi_{iv}>$ formula (\ref{matel}) gives
\begin{eqnarray}
\label{M1iv}
|<iv|\hat O_{11}|0>|^2
&=&-\gamma\frac{\hbar}{2}\Delta_{\bar\L}(\Omega_{iv})/
[2\Omega_{iv}(\Omega^2_{iv}-\Omega_{sc}^2)]
\nonumber\\
&=&\frac{1-\alpha}{8\pi}m\bar\omega^2Q_{00}
\delta^2[E_{iv}^2-2(1+\delta/3)(\hbar\bar\omega)^2]/
[E_{iv}(E^2_{iv}-E_{sc}^2)]\,\mu_N^2.\quad
\end{eqnarray}
For small values of $\delta$ this expression reduces (for
$\alpha=-2$) to
\begin{equation}
\label{isovmat}
|<iv|\hat O_{11}|0>|^2/\mu_N^2\simeq
\frac{3Q_{00}^0}{64\pi}\frac{m\omega_0}{\hbar}
\frac{3\delta^2}
{\sqrt2(1+\delta/2)}\approx
\frac{9}{\sqrt2}
\frac{Q_{00}^0}{64\pi}\frac{m\omega_0}{\hbar}
\,\delta^2.
\end{equation}

Exactly the same results are obtained from the set of equations
for the variables $\bar\R_{2-1},\,\bar\P_{2-1},\,\bar\L_{2-1}$
perturbed by the operator $\hat O_{1-1}$.

Taking into account the relation ${\di Q_{00}^0\frac{m\omega_0}{\hbar}
\simeq\frac{1}{2}\left(\frac{3}{2}A\right)^{4/3}}$, which is usually
\cite{Solov} used to fix the value of the harmonic oscillator
frequency $\omega_0$ , we obtain the following estimate for the
transition probability of the scissors mode:
$$B(M1)\!\uparrow=2\,|<sc|\hat O_{11}|0>|^2=
\frac{(3/2)^{11/6}}{16\pi}A^{4/3}\delta\,\mu_N^2
=0.042A^{4/3}\delta\,\mu_N^2,$$
which practically coincides with the result of
\cite{Lipp}: $B(M1)\!\uparrow=0.043A^{4/3}\delta\,\mu_N^2$,
obtained with  the help of the microscopic approach based on the
evaluation of the sum rules.

\subsection{B(E2)-factors}

Electric transition probabilities can be found exactly in the same way
as the magnetic ones. To calculate the B(E2)-factor it is necessary to
excite the system with the external field operator
\begin{equation}
\label{O2mu}
\hat O_{2\mu'}=er^2Y_{2\mu'}=\beta r^2_{2\mu'},
\quad \hat O_{2\mu'}^{\dagger}=\hat O_{2\mu'}^*=
(-1)^{\mu'}\hat O_{2-\mu'},
\end{equation}
where $\beta=e\sqrt\frac{15}{8\pi}$.
Its Wigner transform is identical to (\ref{O2mu}):
$(O_{2\mu'})_W=\beta r^2_{2\mu'}$.
The matrix element is given by
\begin{equation}
\label{psiO2}
<\psi'|\hat O_{2\mu'}|\psi'>=\beta R_{2\mu'}^{\rm p}=
\frac{1}{2}\beta (R_{2\mu'}-\bar R_{2\mu'}).
\end{equation}
The contribution of $\hat O_{2\mu'}(t)$ to the equation for the Wigner
function is
$$F_{ext}=2\beta\left(F_{\mu'}\exp^{-i\Omega t}+
(-1)^{\mu'}F_{-\mu'}\exp^{i\Omega t}\right)$$
with
$$F_{\mu'}=\sum_{\nu,\sigma}C_{1\nu,1\sigma}^{2\mu'}
r_{\nu}\nabla_{\sigma}^pf^{\rm p}.$$
Integration of $F_{\mu'}$ with the weights
$r_{\lambda\mu}^2~,\,(rp)_{\lambda\mu}$ and $p_{\lambda\mu}^2$
 yields
$$\int d\{\bp,\br\}r_{\lambda\mu}^2F_{\mu'}=0,$$
$$\int d\{\bp,\br\}(rp)_{\lambda\mu}F_{\mu'}=
\sqrt{5(2\lambda+1)}\sum_{k,\pi}
C_{\lambda\mu,2\mu'}^{k,\pi}
\{^{11\lambda}_{k21}\}R_{k\pi}^{\rm p}(eq),$$
$$\int d\{\bp,\br\}p_{\lambda\mu}^2F_{\mu'}=
[1+(-1)^{\lambda}]\sqrt{5(2\lambda+1)}
\sum_{k\pi}C_{\lambda\mu,2\mu'}^{k\pi}
\{^{11\lambda}_{k21}\}
L_{k\pi}^{\rm p}(eq).$$
The external field modifies the set of equations (\ref{quadr}) in the
following way:
\begin{eqnarray}
\label{E2}
\frac{d}{dt}L_{1\mu}^{\rm p}+\cdots&=&
-\beta\sqrt3\left[C_{1\mu,2\mu'}^{2\mu+\mu'}R_{2\mu+\mu'}^{\rm p}(eq)
\exp^{-i\Omega t}
+(-1)^{\mu'}C_{1\mu,2-\mu'}^{2\mu-\mu'}R_{2\mu-\mu'}^{\rm p}(eq)
\exp^{i\Omega t}\right],
\nonumber\\
\frac{d}{dt}L_{2\mu}^{\rm p}-\cdots&=&
\frac{\beta}{\sqrt3}\left[\left(
2\sqrt5C_{2\mu,2\mu'}^{00}R_{00}^{\rm p}(eq)+
\sqrt7C_{2\mu,2\mu'}^{2\mu+\mu'}R_{2\mu+\mu'}^{\rm p}(eq)
\right)\exp^{-i\Omega t}\right.
\nonumber\\
&&\left.+(-1)^{\mu'}\left(
2\sqrt5C_{2\mu,2-\mu'}^{00}R_{00}^{\rm p}(eq)
+\sqrt7C_{2\mu,2-\mu'}^{2\mu-\mu'}R_{2\mu-\mu'}^{\rm p}(eq)
\right)\exp^{i\Omega t}\right].
\end{eqnarray}
The $\mu'=0$ component of the external field does not disturb a
nucleus with axial symmetry. Let us consider the case of
$\mu'=1$ ($\mu'=-1$ gives the same result).
The equations (\ref{E2}) then read
\begin{eqnarray}
\label{inhomE}
&&\dot L_{2\mu}^{\rm p}-\cdots =
\frac{\beta}{3}(Q_{00}^{eq}+\frac{1}{4}Q_{20}^{eq})
\left[\delta_{\mu,-1}\exp^{-i\Omega t}
-\delta_{\mu,1}\exp^{i\Omega t}\right],
\nonumber\\
&&\dot L_{1\mu}^{\rm p}+\cdots=
\frac{\beta}{4}Q_{20}^{eq}\left[\delta_{\mu,-1}\exp^{-i\Omega t}
+\delta_{\mu,1}\exp^{i\Omega t}\right].
\end{eqnarray}

Now, according to formula (\ref{psiO2}), we have to find the
tensors $\bar\R_{21}$ and $\R_{21}$.
The tensor $\bar\R_{21}$ is found by solving the modified
(as in (\ref{inhomE})) set of equations (\ref{scis})
\begin{eqnarray}
\label{scisinhE}
\dot{\bar\R}_{21}-2\bar\L_{21}/m&=&0,
\nonumber\\
\dot{\bar\L}_{21}-\bar\P_{21}/m
+\left[m\,\omega^2+\sqrt{1/6}\chi R_{20}^{eq}
-\sqrt{4/3}\chi_1R_{00}^{eq}\right]\bar\R_{21}&=&
\frac{\beta}{3}(Q_{00}^{eq}+\frac{1}{4}Q_{20}^{eq})
\exp^{i\Omega t},
\nonumber\\
\dot{\bar\P}_{21}
+2[m\omega^2+\sqrt{1/6}\chi_0R_{20}^{eq}]\bar\L_{21}
-\sqrt6\chi_0R_{20}^{eq}\,\bar\L_{11}&=&0,
\nonumber\\
\dot{\bar\L}_{11}
+\sqrt{3/2}\bar{\chi}R_{20}^{eq}\bar\R_{21}&=&
-\frac{\beta}{4}Q_{20}^{eq}\exp^{i\Omega t}.
\end{eqnarray}
It is obvious that the time dependence of all variables
must be $\exp^{i\Omega t}$. The required variable is determined
by the ratio of two determinants
$$\bar\R_{21}=\frac{\Delta_{\bar\R}}{\Delta_{iv}}\exp^{i\Omega t},$$
where $\Delta_{iv}$ is defined by (\ref{harac1}) and
$$\Delta_{\bar\R}=-\frac{\beta}{m}[\frac{2}{3}\Omega^2(Q_{00}^{eq}
+\frac{1}{4}Q_{20}^{eq})+\frac{1}{m}Q_{20}^{eq}
\sqrt{\frac{3}{2}}\chi_0R_{20}^{eq}].$$

The tensor $\R_{21}$ is found by solving the modified
(as in (\ref{inhomE})) set of equations
 (\ref{isosca1}):
\begin{eqnarray}
\dot \R_{21}-2\L_{21}/m&=&0,
\nonumber\\
\dot\L_{21}-\P_{21}/m+
\left[m\,\omega^2+\sqrt{4/3}\chi_0
(R_{20}^{eq}/\sqrt2-R_{00}^{eq})\right]\R_{21}&=&
-\frac{\beta}{3}(Q_{00}^{eq}+\frac{1}{4}Q_{20}^{eq})
\exp^{i\Omega t},
\nonumber\\
\dot\P_{21}
+2[m\omega^2
+\sqrt{1/6}\chi_0R_{20}^{eq}]\L_{21}&=&0,
\nonumber\\
\dot \L_{11}&=&
\frac{\beta}{4}Q_{20}^{eq}\exp^{i\Omega t}.
\label{isoscainhE}
\end{eqnarray}
Again it is obvious that the time dependence of all variables in these
equations must be $\exp^{i\Omega t}$ and the required variable is
determined by the ratio of two determinants
$$\R_{21}=\frac{\Delta_{\R}}{\Delta_{is}}\exp^{i\Omega t},$$
where $\Delta_{is}$ is defined by (\ref{haracis1}) and
$\Delta_{\R}=-\Delta_{\bar\R}$.

The limit (\ref{matel}) is calculated with the help of expression
(\ref{Delta}) for $\Delta_{iv}$ and the analogous expression for
$\Delta_{is}$:
$$\Delta_{is}=(\Omega^2-\Omega_0^2)(\Omega^2-\Omega_{is}^2).$$
In the case $|\psi_a>=|\psi_{sc}>$ we find
\begin{eqnarray}
\label{E2sc}
|<sc|\hat O_{21}|0>|^2&=&-\beta\frac{\hbar}{2}
\Delta_{\bar\R}(\Omega_{sc})/
[(\Omega^2_{sc}-\Omega_{iv}^2)2\Omega_{sc}]
\nonumber\\
&=&\frac{\beta^2\hbar^2}{2m}[\frac{1}{3}E_{sc}^2(Q_{00}^{eq}
+\frac{1}{4}Q_{20}^{eq})+\frac{3\hbar^2}{2m}
\kappa_0(Q_{20}^{eq})^2]/[E_{sc}(E^2_{sc}-E_{iv}^2)]
\nonumber\\
&=&\frac{e^2\hbar^2}{m}\frac{5}{16\pi}Q_{00}[
2(\hbar\bar\omega\delta)^2-(1+\frac{\delta}{3})E_{sc}^2
]/[E_{sc}(E^2_{iv}-E_{sc}^2)].
\end{eqnarray}
For small $\delta$ (and $\alpha=-2$)
$$|<sc|\hat O_{21}|0>|^2\simeq
\frac{e^2\hbar}{m\omega_0}\frac{5}{128\sqrt6 \pi}Q_{00}^0
\frac{\delta}
{1-\delta/6}\approx
\frac{e^2\hbar}{m\omega_0}\frac{5}{128\sqrt6 \pi}Q_{00}^0
\delta.$$

In the case $|\psi_a>=|\psi_{iv}>$ formula (\ref{matel}) gives
\begin{eqnarray}
\label{E2iv}
|<iv|\hat O_{21}|0>|^2&=&-\beta\frac{\hbar}{2}
\Delta_{\bar\R}(\Omega_{iv})/
[2\Omega_{iv}(\Omega^2_{iv}-\Omega_{sc}^2)]
\nonumber\\
&=&\frac{\beta^2\hbar^2}{2m}[\frac{1}{3}E_{iv}^2(Q_{00}^{eq}
+\frac{1}{4}Q_{20}^{eq})+\frac{3\hbar^2}{2m}
\kappa_0(Q_{20}^{eq})^2]
/[E_{iv}(E^2_{iv}-E_{sc}^2)]
\nonumber\\
&=&\frac{e^2\hbar^2}{m}\frac{5}{16\pi}Q_{00}
[(1+\frac{\delta}{3})E_{iv}^2-2(\hbar\bar\omega\delta)^2]
/[E_{iv}(E^2_{iv}-E_{sc}^2)].
\end{eqnarray}
For small values of $\delta$ (and $\alpha=-2$)
this expression reduces to
$$|<iv|\hat O_{21}|0>|^2\simeq
\frac{e^2\hbar}{m\omega_0}\frac{5}{32\sqrt2\pi}Q_{00}^0
\frac{1+\frac{2}{3}\delta}
{1+\delta/6}\approx
\frac{e^2\hbar}{m\omega_0}\frac{5}{32\sqrt2\pi}Q_{00}^0.$$

In the case $|\psi_a>=|\psi_{is}>$ formula (\ref{matel}) gives
\begin{eqnarray}
\label{E2is}
|<is|\hat O_{21}|0>|^2&=&\!-\beta\frac{\hbar}{2}
\Delta_{\R}(\Omega_{is})/
[2\Omega_{is}(\Omega^2_{is}-\Omega_{0}^2)]
\nonumber\\
&=&\frac{\beta^2\hbar^2}{2m}[\frac{1}{3}E_{is}^2(Q_{00}^{eq}
+\frac{1}{4}Q_{20}^{eq})+\frac{3\hbar^2}{2m}
\kappa_0(Q_{20}^{eq})^2]/[E_{is}(E^2_{is}-\hbar^2\Omega_0^2)]
\nonumber\\
&=&\frac{e^2\hbar^2}{m}\frac{5}{16\pi}Q_{00}
[(1+\frac{\delta}{3})E_{is}^2
-2(\hbar\bar\omega\delta)^2]/[E_{is}]^3.
\end{eqnarray}
For small values of $\delta$ this expression reduces to
$$|<is|\hat O_{21}|0>|^2 \simeq
\frac{e^2\hbar}{m\omega_0}\frac{5}{16\sqrt2\pi}Q_{00}^0(1+\delta/3).$$

Formula (\ref{matel}) allows one to calculate the matrix element
$|<\psi_a|\hat O|\psi_0>|^2$ also in the case when
$|\psi_a>=|\Omega_0>$, i.e., for the rotational state corresponding
to the trivial solution of (\ref{haracis2}):
\begin{eqnarray}
\label{E2is0}
|<\Omega_0|\hat O_{21}|0>|^2&=&\!-\beta\frac{\hbar}{2}
\Delta_{\R}(\Omega_0)/
[2\Omega_0(\Omega^2_{0}-\Omega_{is}^2)]
\nonumber\\
&=&\!\frac{\beta^2\hbar^2}{2m}[\frac{1}{3}\hbar^2\Omega_0^2
(Q_{00}^{eq}+\frac{1}{4}Q_{20}^{eq})+\frac{3\hbar^2}{2m}
\kappa_0(Q_{20}^{eq})^2]/[\hbar\Omega_0(\hbar^2\Omega_0^2-E^2_{is})]
\nonumber\\
&=&\frac{e^2\hbar^2}{m}\frac{5}{8\pi}
Q_{00}\delta^2/[\hbar\Omega_0 2(1+\delta/3)].
\end{eqnarray}
The value of this matrix element is obviously
infinite due to the zero value of $\Omega_0$. However, below
this expression
will be useful to calculate the energy weighted sum rule.

\section{Sum rules}

\subsection{Magnetic case}

The magnetic dipole operator (\ref{Magnet1}) is not hermitian.
By definition it is a linear combination of hermitian operators
(components of the angular momentum)
$$\hat O_{11}=-\frac{i}{2}\gamma (\hat I_x+i\hat I_y),\quad
\hat O_{1-1}=\frac{i}{2}\gamma (\hat I_x-i\hat I_y).$$
This fact allows one to derive several useful relations:
$$[\hat O_{11},[H,\hat O_{1-1}]]=
\frac{\gamma^2}{4}([\hat I_x,[H,\hat I_x]]+
[\hat I_y,[H,\hat I_y]]),$$
\begin{eqnarray}
<0|\hat O_{11}|\nu><\nu|\hat O_{1-1}|0>&=&
\frac{\gamma^2}{2}(|<\nu|\hat I_x|0>|^2+|<\nu|\hat I_y|0>|^2)
\nonumber\\
&=&
-(|<\nu|\hat O_{11}|0>|^2+|<\nu|\hat O_{1-1}|0>|^2).
\nonumber
\end{eqnarray}
Using these formulae and the standard sum rule for a hermitian
operator \cite{BM}
$$\sum_{\nu}(E_{\nu}-E_0)
|<\nu|\hat I_i|0>|^2
=\frac{1}{2}<0|[\hat I_i,[H,\hat I_i]]|0>,$$
one immediately obtains the sum rule for $\hat O_{1\pm1}$:
\begin{eqnarray}
\label{sumrule}
\sum_{\nu}(E_{\nu}-E_0)
(|<\nu|\hat O_{11}|0>|^2+|<\nu|\hat O_{1-1}|0>|^2)
=
-<0|[\hat O_{11},[H,\hat O_{1-1}]]|0>.
\end{eqnarray}
 It can also be calculated in a more direct way:
\begin{eqnarray}
\label{dsumrule}
&&<0|[\hat O_{11},[H,\hat O_{1-1}]]|0>=
\\
\nonumber
&&=\sum_{\nu}(E_{\nu}-E_0)
(<0|\hat O_{11}|\nu><\nu|\hat O_{1-1}|0>+
<0|\hat O_{1-1}|\nu><\nu|\hat O_{11}|0>)
\\
\nonumber
&&=\sum_{\nu}(E_{\nu}-E_0)
(<0|\hat O_{11}|\nu><0|\hat O_{1-1}^{\dagger}|\nu>^*+
<0|\hat O_{1-1}|\nu><0|\hat O_{11}^{\dagger}|\nu>^*).
\end{eqnarray}
Using here the hermitian conjugation properties (\ref{Magnet1}) of
the operator $\hat O_{1\mu}$, one reproduces formula (\ref{sumrule}).

The double commutator is calculated with the help of
(\ref{Ham}) and (\ref{Magnet1}):
\begin{equation}
\label{commut}
[\hat O_{1\phi},[H,\hat O_{1\phi'}]]=
\frac{15}{2\pi}\bar\chi\sum_i^Z
\sum_j^N\sum_{\nu,\sigma,\epsilon}(-1)^{\nu}C_{2\nu,2\sigma}^{1\phi}
C_{2-\nu,2\epsilon}^{1\phi'}r_{2\epsilon}^2(i)r_{2\sigma}^2(j)\mu_N^2.
\end{equation}
Taking into account axial symmetry, one finds the ground state
matrix element of (\ref{commut}) (in Hartree-Fock approximation)
$$<0|[\hat O_{1\phi},[H,\hat O_{1\phi'}]]|0>/\mu_N^2=
\frac{15}{2\pi}\bar\chi
\sum_{\nu}(-1)^{\nu}C_{2\nu,2 0}^{1\phi}
C_{2-\nu,2 0}^{1\phi'}R_{2 0}^{\rm p}R_{2 0}^{\rm n}=
\frac{15}{8\pi}\delta_{\phi',-\phi}\bar\chi
(C_{2\phi,2 0}^{1\phi}R_{2 0}^{eq})^2.$$
It is obvious that this expression is different from zero only
for $\phi=\pm 1$. Hence, the final expression for the right-hand side
of (\ref{sumrule}) is
\begin{equation}
\label{Rfin}
<0|[\hat O_{11},[H,\hat O_{1-1}]]|0>=
\frac{9}{16\pi}\bar\chi (R_{2 0}^{eq})^2\mu_N^2
=-\frac{1-\alpha}{4\pi}Q_{00}m\bar\omega^2\delta^2\mu_N^2\equiv
-(1-\alpha)\Sigma_0,
\end{equation}
where, for the sake of convenience, the notation
$\di \Sigma_0=\frac{m\bar\omega^2}{4\pi}Q_{00}\delta^2\mu_N^2$ is
introduced.
The left-hand side of (\ref{sumrule}) is calculated trivially
by multiplying the right-hand side of (\ref{scimat}) by $E_{sc}$ and
adding it to the second line of (\ref{M1iv}) multiplied by $E_{iv}$:
\begin{eqnarray}
\Sigma_{tot}&=&
\sum_{\nu}(E_{\nu}-E_0)
\left(|<\nu|\hat O_{11}|0>|^2+|<\nu|\hat O_{1-1}|0>|^2
\right)
\nonumber\\
&=&2\left(E_{sc}|<sc|\hat O_{11}|0>|^2+E_{iv}|<iv|\hat O_{11}|0>|^2
\right)
\nonumber\\
&=&\Sigma_{sc}+\Sigma_{iv}=(1-\alpha)\Sigma_0\,,
\label{LHS}
\end{eqnarray}
where
\begin{equation}
\label{sumscex}
\Sigma_{sc}=\frac{[E_{sc}^2-2(1+\delta/3)(\hbar\bar\omega)^2]}
{(E^2_{sc}-E_{iv}^2)}(1-\alpha)\Sigma_0
\end{equation}
and
\begin{equation}
\label{sumiv}
\Sigma_{iv}=\frac{[E_{iv}^2-2(1+\delta/3)(\hbar\bar\omega)^2]}
{(E^2_{iv}-E_{sc}^2)}(1-\alpha)\Sigma_0\,.
\end{equation}
So, one sees that the sum rule (\ref{sumrule}) is fulfilled.

It is useful to estimate the contribution to the sum rule
from the scissors mode in the small deformation
limit. First of
all, with the help of formula (\ref{Omeg2}), one evaluates the
difference
$$E_{iv}^2-E_{sc}^2\simeq 2(\hbar \bar\omega)^2(2-\alpha)
(1+\frac{\delta}{3}).$$
With this the contribution of the scissors mode is calculated quite
easily:
\begin{eqnarray}
&&\Sigma_{sc}=(1-\alpha)\Sigma_0\,
\frac{[2(1+\delta/3)(\hbar\bar\omega)^2-E_{sc}^2]}
{(E^2_{iv}-E_{sc}^2)}\simeq
\Sigma_0\,\frac{1-\alpha}{2-\alpha}.
\label{sumscis}
\end{eqnarray}
Neglecting the $\delta^3$-term and taking
$\hbar\omega_0=41/A^{1/3}$MeV (such value was used in the papers
\cite{Bes,HamAb}) we reproduce the result of these papers
\begin{equation}
\label{sumscis1}
\Sigma_{sc}\simeq
\frac{41}{8\pi}\left( \frac{3}{2}\right)^{\frac{4}{3}}A\delta^2
\,\frac{1-\alpha}{2-\alpha}\mu_N^2{\rm MeV}
=1.4\,\frac{2-2\alpha}{2-\alpha}\delta^2A \mu_N^2{\rm MeV}.
\end{equation}
For further discussion see section 7.5.

\subsection{Electric case}

The sum rule for $\hat O_{2\pm1}$ can easily be obtained by
replacing in formula (\ref{dsumrule}) the operators
$\hat O_{1\pm1}$ by the operators $\hat O_{2\pm1}$ and
using the hermitian conjugation properties
(\ref{O2mu}) of the operator $\hat O_{2\mu}$:
\begin{eqnarray}
\label{sumruleO2}
\sum_{\nu}(E_{\nu}-E_0)
(|<\nu|\hat O_{21}|0>|^2+|<\nu|\hat O_{2-1}|0>|^2)
=
-<0|[\hat O_{21},[H,\hat O_{2-1}]]|0>.
\end{eqnarray}
The double commutator is calculated with the help of
(\ref{Ham}) and (\ref{O2mu}):
\begin{equation}
\label{commut2}
[\hat O_{2\phi},[H,\hat O_{2\phi'}]]=
-20\beta^2\frac{\hbar^2}{m}
\sum_i^Z\sum_{\lambda,\sigma}C_{2\phi,2\phi'}^{\lambda\sigma}
\{^{112}_{\lambda21}\}r_{\lambda\sigma}^2(i).
\end{equation}
Taking into account axial symmetry, one finds the ground state
matrix element of (\ref{commut2}):
\begin{eqnarray}
<0|[\hat O_{2\phi},[H,\hat O_{2\phi'}]]|0>&=&
-20\beta^2\frac{\hbar^2}{m}
\delta_{\phi,-\phi'}\sum_{\lambda=0,2}C_{2\phi,2 -\phi}^{\lambda 0}
\{^{112}_{\lambda21}\}R_{\lambda0}^{\rm p}
\nonumber\\
&=&
-2\beta^2\frac{\hbar^2}{m}
\delta_{\phi,-\phi'}\left((-1)^{\phi}\frac{2}{\sqrt3}R_{00}^{\rm p}+
\frac{1}{\sqrt6}R_{20}^{\rm p}\right).
\label{commut21}
\end{eqnarray}
Taking here $\phi=1$ we obtain the final expression for the
right-hand side of (\ref{sumruleO2})
$$<0|[\hat O_{21},[H,\hat O_{2-1}]]|0>=
-2\beta^2\frac{\hbar^2}{m}
\left(\frac{2}{3}Q_{00}^{\rm p}+
\frac{1}{6}Q_{20}^{\rm p}\right)=-e^2\frac{\hbar^2}{m}\frac{5}{4\pi}
Q_{00}(1+\delta/3).$$

The left-hand side of (\ref{sumruleO2}) is calculated by summing
expressions (\ref{E2sc}), (\ref{E2iv}), (\ref{E2is}) and
(\ref{E2is0}) multiplied by the respective energies. It is convenient
to calculate
the isovector and isoscalar contributions separately. The contribution
of the isovector modes is
\begin{eqnarray}
2\left(E_{sc}|<sc|\hat O_{21}|0>|^2+E_{iv}|<iv|\hat O_{21}|0>|^2
\right)&=&\frac{\beta^2\hbar^2}{3m}(Q_{00}+\frac{1}{4}Q_{20})
\nonumber\\
&=&e^2\frac{\hbar^2}{m}\frac{5}{8\pi}Q_{00}(1+\delta/3).
\label{LHSiv}
\end{eqnarray}
Exactly the same result is obtained for isoscalar modes:
\begin{eqnarray}
2\left(\hbar\Omega_0|<\Omega_0|\hat O_{21}|0>|^2
+E_{is}|<is|\hat O_{21}|0>|^2
\right)
=e^2\frac{\hbar^2}{m}\frac{5}{8\pi}Q_{00}(1+\delta/3).
\label{LHSis}
\end{eqnarray}
 Hence the
sum rule (\ref{sumruleO2}) is fulfilled.

It is interesting to compare the contributions of the scissors mode
and the rotational mode. The scissors mode (for small $\delta$) yields:
\begin{eqnarray}
2E_{sc}|<sc|\hat O_{21}|0>|^2
\simeq\frac{5}{128\pi}e^2\frac{\hbar^2}{m}Q_{00}\delta^2.
\label{sumsc}
\end{eqnarray}
The rotational mode yields:
\begin{eqnarray}
2\hbar\Omega_0|<\Omega_0|\hat O_{21}|0>|^2=
\frac{5}{8\pi}e^2\frac{\hbar^2}{m}Q_{00}\frac{\delta^2}{1+\delta/3}.
\label{sumnonv}
\end{eqnarray}
It is seen that the contribution of the rotational mode is
approximately 16 times larger than the one of the scissors mode.
This is a very significant number demonstrating the importance of
excluding the spurious state from the theoretical results. Indeed,
to describe correctly such a subtle phenomenon as the scissors mode,
it is compulsory to eliminate the errors from spurious motion whose
value can be an order of magnitude larger than the phenomenon under
consideration.

\section{Discussion}

\subsection{Hierarchy of variables}

 Let us analyze carefully the set of equations (\ref{scis}).
It contains a minimal set of variables
required to describe the discussed phenomenon - scissors mode.
 The information of the first equation
is more or less trivial: the tensor $\bar \L_{2\mu}$ is just the time
derivative of the quadrupole moment $\bar \R_{2\mu}$. Thus, one can
say that equations (\ref{scis}) describe the coupled dynamics of the
angular momentum $\bar \L_{11}(t)$, the quadrupole
moment $\bar \R_{21}(t)$ and the quadrupole kinetic energy tensor
$\bar \P_{21}(t)$. And, what is of principal importance, the angular
momentum does not play the key role in this ensemble. It is possible
to neglect this variable without any serious consequence for the rest
of equations, which will in such a case describe the isovector GQR
(see formulae (\ref{rotzero},\,\ref{IVGQR})).

 The variables $\bar \R_{21}(t)$ and $\bar \P_{21}(t)$ are of
considerably more fundamental character. It is obvious that
one cannot neglect the quadrupole moment
$\bar \R_{21}(t)$ because it is the basis of the ensemble
and the whole problem loses any physical meaning without
$\bar \R_{21}(t)$.
The kinetic energy tensor
$\bar \P_{21}(t)$ is responsible for the Fermi surface deformation and
must be taken into account to correctly describe the elastic
properties of nuclei \cite{Bertsch} and, as a result, to get the
correct value of the GQR energy.
 Thus, one arrives at the conclusion: it is
impossible to construct a reasonable model of the scissors mode
with only one pair of variables, the angular momentum
$\bar \L_{11}(t)$ and its canonically conjugate variable. The
scissors mode can not exist independently, on its own, without being
coupled to the IVGQR (see below). This
conclusion is in absolute contradiction with the original idea
of N. Lo Iudice and F. Palumbo \cite{Lo} and especially with their
two rotor model (TRM) underlying this idea.

\subsection{Rotation due to vibration}

Rather soon it was understood \cite{Lipp,Hilt84} that the
rotational motion must be accompanied by the isovector quadrupole
vibration
(shear mode), the second kind of motion being a small admixture to
the first one. However, this statement, being true in essence, can be
misleading to capture the phenomenon.
Indeed, one can easily
come to the conclusion that the rotational motion is the principal
constituent of the phenomenon, and the vibrational motion is
accessory and can be neglected to simplify the
description of the problem, if one is interested in qualitative
results only. It is easy to see, however, that the real situation is
exactly the inverse! Our analysis of the set of equations
(\ref{scis}) has shown
that the rotational motion (the variable $\bar \L_{11}(t)$) exists
only due to the vibrational one (the variables $\bar \R_{21}(t),\,
\bar \P_{21}(t)$). If one wants to observe the scissors mode,
one has to excite the IVGQR simultaneously. The
IVGQR can exist without the
scissors mode, but the scissors mode cannot exist without the IVGQR!
Neglecting the coupling to the quadrupole deformation in the last of
equations (\ref{scis}) would induce a full free counter rotational
motion of neutrons versus protons!

The only characteristic, in which the rotational motion exceeds the
vibrational one, is the value of the B(M1)-factor (see, however, the
section 8). The ratio
$\di \frac{B(M1)_{iv}}{B(M1)_{sc}}\simeq\frac{3\sqrt3}{4}\delta$ is of
the same order of magnitude as the coefficient serving to measure
the contribution of the vibrational motion to the scissors mode:
$\di \eta=\frac{\omega_y-\omega_z}{\omega_y+\omega_z}\simeq
\frac{\delta}{2}$ (see papers \cite{Hilt84,Hilt92}) and
$\di \alpha=\delta/(1+\frac{1}{2}\Omega^2_{is}/\Omega^2_D)$ in
paper \cite{Lipp} ($\Omega_D$ is a frequency of a giant dipole
resonance).

\subsection{Fermi surface deformation}

As a matter of fact, the most important ingredient to the scissors
mode is the Fermi surface deformation. It can be understood by
analyzing formulae for the eigenfrequences of all three modes. Neglecting
the variable $\P_{21}$ in (\ref{isosca1}) we find the following
expression for the frequency of ISGQR:
$$\Omega_{is}^2=\frac{2}{m}[m\omega^2
+\frac{2}{\sqrt3}\chi_0(R_{20}^{eq}/\sqrt2-R_{00}^{eq})]
=2[\omega^2+4\frac{\kappa_0}{m}Q_{00}(1+\frac{2}{3}\delta)].$$
For self-consistent value of the strength constant $\di \kappa_0=
-\frac{m\bar\omega^2}{4Q_{00}}$ one obtains
$\Omega^2_{is}=0.$ This result is quite natural, because the pure
geometric distortion corresponding to $\R_{21}$ can be produced by
the proper rotation of the nucleus, without any disturbance of its
internal structure.
Neglecting the variable $\bar \P_{21}(t)$ in (\ref{scis}) we find that
the frequency of IVGQR (being determined mainly by the neutron-proton
interaction) is changed not so drastically:
$$\Omega_{iv}^2=2\bar\omega^2(1-\alpha)(1+\delta/3).$$
Comparing this formula (for $\alpha=-2$ ) with (\ref{Eniv}) one sees,
that $\Omega_{iv}^2\simeq8\omega^2_0$ changes to
 $\Omega_{iv}^2\simeq6\omega^2_0$. One should recall that also for the
Isovector Giant Dipole Resonance
the distortion of Fermi sphere plays only a minor role.

 It is also easy to see
that omitting $\bar \P_{21}(t)$ in (\ref{scis}), one
obtains zero energy for the scissors mode independent of the
strength of the residual interaction.

Thus, the nuclear elasticity
discovered by G.F.Bertsch \cite{Bertsch} is the single origin for the
restoring force of the scissors mode and also for the ISGQR
in our simple Hamiltonian of a harmonic oscillator with
Q-Q residual interaction.
 So one can conclude that this mode is in its essence a pure quantum
mechanical phenomenon. This agrees with the conclusion of the papers
\cite{Lipp9,ZawSp}: classically (i.e., without Fermi surface
deformation) the scissors mode is a zero energy mode.

\subsection{Deformed oscillator and isovector Q-Q interaction}

It is known that the deformed harmonic oscillator Hamiltonian can be
obtained in a Hartree approximation "by making the assumption that the
isoscalar part of the Q-Q force builds the one-body container well"
\cite{Hilt92}. Thus, neglecting the isovector part of the Q-Q
residual interaction, i.e. assuming $\kappa_1=0 \to
\kappa=\bar \kappa=\kappa_0$, we have to reproduce the known results
in the deformed harmonic oscillator model. The formulas of other
authors are obtained, as a rule, in the small $\delta$ limit. For the
sake of convenient comparison just the same approximation is used
here. We have for the energies
$$\Omega_{iv}\simeq 2\omega_0,\quad
\Omega_{sc}\simeq\omega_0\delta,$$
which coincides with the results of \cite{Hilt92} and \cite{Garrid}.
The formulae for magnetic transition probabilities are
$$B(M1)_{sc}\!\uparrow\simeq \frac{1}{8\pi}
\frac{m\omega_0}{\hbar}Q_{00}^0\delta\,\mu_N^2,\quad
B(M1)_{iv}\!\uparrow\simeq \frac{1}{16\pi}
\frac{m\omega_0}{\hbar}Q_{00}^0\delta^2\mu_N^2,\quad
\frac{B(M1)_{iv}}{B(M1)_{sc}}=\frac{\delta}{2},$$
which coincide with the results of \cite{Zaw}. Possibly they coincide
also with that of \cite{Garrid} (as a matter of fact,
their values are twice larger, but we suppose it is a misprint).
For electric transition probabilities we find
$$B(E2)_{sc}\!\uparrow\simeq \frac{5}{64\pi}
\frac{e^2\hbar}{m\omega_0}Q_{00}^0\delta,\quad
B(E2)_{iv}\!\uparrow\simeq \frac{5}{32\pi}
\frac{e^2\hbar}{m\omega_0}Q_{00}^0,\quad
\frac{B(E2)_{sc}}{B(E2)_{iv}}=\frac{\delta}{2},$$
in perfect agreement with \cite{Zaw}.
The ratios of different characteristics, calculated with and without
an isovector Q-Q interaction, are
$$\frac{\Omega_{sc}}{\Omega_{sc}(\kappa_1=0)}=\sqrt{\frac{3}{2}},
\quad\di \frac{\Omega_{iv}}{\Omega_{iv}(\kappa_1=0)}=\sqrt2,$$
$$\frac{B(M1)_{sc}}{B(M1)_{sc}(\kappa_1=0)}=
\sqrt{\frac{3}{2}},\quad
\frac{B(M1)_{iv}}{B(M1)_{iv}(\kappa_1=0)}=
\frac{9}{2\sqrt2},$$
$$\frac{B(E2)_{sc}}{B(E2)_{sc}(\kappa_1=0)}=
\frac{1}{2\sqrt6},\quad
\frac{B(E2)_{iv}}{B(E2)_{iv}(\kappa_1=0)}=
\frac{1}{\sqrt2}.$$
As we can see, the inclusion of an isovector Q-Q interaction increases the
energies and B(M1)-factors of the scissors mode and IVGQR and
decreases their B(E2)-factors. It is also necessary to emphasize the
following important result: the ratio
$$RM\equiv B(M1)_{iv}(\kappa_1=0)/
B(M1)_{sc}(\kappa_1=0)$$
coincides exactly with the "admixture" coefficient $\eta$ introduced
by Hilton \cite{Hilt84}, what supports our idea that $RM$ can serve as
a measure for an "admixture". It is seen that taking into account
the long range correlations (isovector Q-Q forces), one increases $RM$
by the factor $3\sqrt3/2$.
And finally, the ratio $\Omega_{sc}/\Omega_{sc}(\kappa_1=0)$
is quite close to the number $\sqrt{(1+0.66)}$ found in \cite{Hilt92}.

\subsection{Sum rule}
Our discussion of the magnetic sum rules will be based on table 4
of the review of Zawischa \cite{Zaw} where the results of
different models and approaches for $^{164}$Dy are listed. For the
sake of a convenient comparison it is reproduced here
(together with its legend) as table 1.

\noindent{\bf Table 1.} The energy-weighted orbital M1 sum rule
$\sum E_x B(M1_{orb})\uparrow$ (in units of $\mu_N^2$MeV) in
different models evaluated for $^{164}$Dy as an example, compared
with the values obtained from the expressions given by Lipparini and
Stringari \cite{Lipp}. The total sum rule strength of the model and
the part exhausted by the low-energy mode are given. The schematic
RPA results are from Bes and Broglia \cite{Bes}, for the RPA with
Migdal force entry the data of Zawischa and Speth \cite{ZawSp} are
used. In lines 4 and 5 the deformation parameter $\delta_{osc}=0.258$
has been taken.\\
\vspace{-0.8cm}
\begin{center}
\begin{tabular}{lll}
\hline
                             & Total & Low energy  \\
\hline
TRRM                         & 140.8 & -           \\
Classical fluids             & 140.8 & 0           \\
NFD model                    & 141.6 & 37.8        \\
Deformed harmonic oscillator & 31.0  & 15.4        \\
Schematic RPA                & $>$70 & 24.4        \\
RPA with Migdal force        & 108.2 & 40.6        \\
IBM-2 (with $E_x=3$ MeV)     & 12.2  & 12.2        \\
Lipparini and Stringari \cite{Lipp}& 145.4 & 35.9  \\
\hline
\end{tabular}
\end{center}
\vspace{0.5cm}

To complete the legend of this table, it should be said that the
results of the lines 1, 2, 3, 4 and 7 are calculated by Zawischa
\cite{Zaw}, who used $\hbar\omega_0=46.5/A^{1/3} {\rm MeV}$ (what
corresponds to $r_0=1.13$ fm) and $\delta=0.302$ (except the fourth
line). Taking the same
values of parameters one obtains $\Sigma_0=42.4\mu_N^2$MeV.
 In the case of $\alpha=-2$ one finds from (\ref{LHS})
$$\Sigma_{tot}=3\Sigma_0=127.3 \mu_N^2{\rm MeV}.$$
It is seen from table 1 that this number does not contradict the
"Schematic RPA" and is in qualitative agreement with the lines "TRRM"
(Two Rigid Rotors Model), "Classical fluids", "NFD model" (Nuclear
Fluid Dynamics), "Lipparini and Stringari \cite{Lipp}" and "RPA with
Migdal force", being exactly in between the last two results. We have
to note that the result of the small $\delta$ approximation
($\bar\omega \to \omega_0,\, Q_{00} \to Q_{00}^0$)
$$\Sigma_{tot}=3\Sigma_0=142.6 \mu_N^2{\rm MeV}$$
is in excellent agreement with "NFD model" and "Lipparini and
Stringari \cite{Lipp}" lines whose values were obtained in the small
approximation also. Such agreement
in the last case is especially surprising if one takes into account
that the $M1$ sum rule is model dependent, being determined by the
neutron-proton interaction, which is quite different in
the two papers: quadrupole-quadrupole residual interaction in our case
and that of the Skyrme type in \cite{Lipp}. This fact confirms that the
model Hamiltonian used in this work is realistic enough.
The exact contribution of the scissors mode to the sum
rule (formula (\ref{sumscex})) is
$$\Sigma_{sc}=30.9 \mu_N^2{\rm MeV}.$$
 The result of the small $\delta$ approximation (formula
(\ref{sumscis1}) with $\hbar\omega_0=46.5/A^{1/3} {\rm MeV}$)
$$\Sigma_{sc}=35.6 \mu_N^2{\rm MeV}$$
is rather close to the exact number, being in
good agreement with "NFD model" and
"Lipparini and Stringari \cite{Lipp}" lines.
It is worth noting that the
 ratio $\Sigma_{sc}/\Sigma_{iv}\simeq1/3$ is not so far from
the value $\Sigma_{sc}/\Sigma_{high}\simeq1/4$ predicted in \cite{Enders}
on the basis of some theoretical analysis of experimental data
(their notation ``high'' means high energy scissors mode).

In the case of $\alpha=0$, which corresponds to the deformed harmonic
oscillator, one has (for $\delta=0.258$, as in \cite{Zaw})
$$\Sigma_{tot}=31.2 \mu_N^2{\rm MeV},\quad
\Sigma_{sc}=\Sigma_{iv}=15.6 \mu_N^2{\rm MeV}$$
reproducing the numbers of the line
"Deformed harmonic oscillator" in table 1.

Concluding this subsection we have to say that most of the general
observations found earlier \cite{Zaw} (such as, for example,
$\Sigma_{sc}/\Sigma_{iv}\simeq 1/3,\,\Sigma_{sc}\sim \delta^2$) are
confirmed by our investigation.

\subsection{Discussion of other approaches to the scissors mode}

After having, as we think, clearly worked out the physics of the low
lying scissors mode and its interweaving with the IVGQR, one may ask
the question about the status of other approaches. As already
mentioned all other RPA-type approaches \cite{Hamam,Suzuki,Bes,HamAb}
have from the numerical point of view the same status as our approach
(but this concerns mostly only the limit of small deformations). We
also have pointed out that the original model of counter
rotating rigid rotors
\cite{Lo,Lo2000} does, in our opinion, not at all grasp the salient
features of the scissors mode. However, the description of the
scissors mode also has been attempted with other quite different
approaches like IBM (or IBA), shell model calculations, etc.
We think that the conclusions concerning IBM (IBA) are very well
formulated in the review by Zawischa \cite{Zaw}.
Therefore, to demonstrate our point of view we will simply
comment several citations from \cite{Zaw}.

i) "The original aim of the IBA was the microscopic explanation of
vibrations, rotations and transitions between the two bosons.
Rotational invariance is maintained throughout. Starting from a
microscopic Hamiltonian, the configuration space is drastically
truncated by dealing only with nucleons (or holes) in the valence
shell, assuming an inert core, and approximating correlated pairs of
valence nucleons by bosons \cite{Otsuka1}. (Mostly monopole and
quadrupole bosons are considered, for special purposes bosons with
angular momentum different from 0 or 2 are needed too.)...

The bosons are treated as elementary units, but the internal
structure of the fermion pairs they represent, reflects itself in the
parameters of the model Hamiltonian which are fitted to the low energy
spectra and transition probabilities \cite{Otsuka2}.
 By construction, the model is only suitable to describe low-energy
states. To describe the giant dipole resonance, $p$ and $f$ bosons
have been included (\cite{Heyde} and references therein.) The giant
quadrupole resonances are outside of its scope, as they involve
$\Delta N_{osc}=2$ excitations ($N_{osc}$ is the oscillator quantum
number). We have seen that (in the semiclassical model, in the
deformed harmonic oscillator model and in microscopic RPA) a
considerable amount of strength is in a high-energy mode--the
$|K|=1$ component of the isovector giant quadrupole resonance. Due to
truncation of the configuration space, this strength is missing in
the IBM-2 sum rule. We have also seen that (in microscopic RPA) a
considerable part of the M1 strength resides in the region between 4
and 10 MeV, in two-quasiparticle type excitations--all these are not
included in the model space of the IBA....
Thus, the IBA sum rule by the basic assumptions of the model comprises
only a small part of the full sum rule. Therefore, it is not so well
suited to assess the collectivity of a state."

ii) "As already mentioned, among the predictions, the IBA was closest
to the strengths to be detected. The reason is, of course, that the
model parameters are extracted from low-energy data and the model is
best adapted just to the energy range up to a few MeV where the M1
states have been found...
The naive assumption of bare orbital g-factors already gives
a good prediction of the low-energy strength. Adjustment of the model
parameters to the low-energy vibrational and rotational states
\cite{Otsuka2} finally has the effect that the average energy is also
well reproduced."

iii) "In order to find out the physical interpretation of the lowest
$K=1$ mixed symmetry state of the IBM-2, the (semi)classical limit of
the IBM-2 Hamiltonian has been investigated by Balantekin and Barrett
\cite{Balant}, Bijker \cite{Bijker} and Walet \cite{Walet}, yielding
a Hamiltonian similar to that of TRRM, the potential being a function
of the angle $\theta$ between the symmetry axes of protons and
neutrons: $\,V(\theta)=\lambda\theta^2$."

With respect to iii) we must note, that
deriving the classical limit, the vibrational degrees of freedom
have been neglected to simplify the derivation. Not surprisingly
they obtained the TRRM Hamiltonian as a result for which we have
already given our opinion above.

 Generally speaking these above citations give exaustive
characteristic of the status of IBM (IBA) calculations of the
scissors mode: they are able to
give correct values for the energy and strength of the scissors mode
but they do not explain the real physics of the phenomenon.

 The situation with shell model calculations is rather complicated.
There are the well known difficulties to treat heavy nuclei because
of the huge dimension of matrices. Therefore the calculations are
usually made for very light nuclei. Even there one must divide them
into two groupes. There are qualitative estimations with truncated
basis ($\Delta N_{osc}=0$, see for example \cite{Zamick,Poves}).
Naturally, those calculations suffer from the same drawbacks as the
IBM calculations and then the same comment given above apply.
There are also 'realistic' calculations (for $^8$Be and $^{10}$Be)
with an extended
$(\Delta N_{osc}=0)+(\Delta N_{osc}=2)$ basis \cite{Fayache}.
 In principle they have the same status as RPA calculations.
Still one can ask the question whether it makes sense to talk
    about scissors mode in such light nuclei.

 We also would like to comment on the neutron-proton deformation (NPD)
model. We again cite \cite{Zaw}:
"The Bohr-Mottelson model has been generalized to isovector degrees
of freedom leading to the neutron-proton deformation (NPD) model...
Rohozinski and Greiner \cite{Rohoz} applied the NPD model to the
orbital magnetic dipole excitations....

The TRRM also has been changed by its authors to the TRM, relaxing
the condition  of rigid rotation and replacing the rigid-body moment
of inertia by a smaller one obtained from some model or a
phenomenological one \cite{LoRi}." If their parameters
"...are adjusted to the data,
the TRM and NPD model will coincide, even though in their original
assumptions--rigid rotation in the one, irrotational flow in the
other--the models are contradictory."

 Therefore we agree with the conclusion of Zawischa \cite{Zaw} that
the NPD model can not describe the scissors mode. The reasons are
also especially well born
out in terms of our method: the collective variables in their model
are only $\bar\R_{2\mu}$ and $\bar\L_{2\mu}$ , there are no rotational
degrees of freedom ($\bar\L_{1\nu}$) and there is no Fermi surface
deformation ($\bar\P_{2\mu}$). As we know from an earlier discussion,
only the isovector giant resonance survives in such conditions (see
section 7.3). Therefore, the $\delta^2$-dependence of $B(M1)$-factor
which the NPD model predicts and which is in principle in agreement
with experiment, must actually be interpreted as the
$\delta^2$-dependence of $B(M1)$ for the IVGQR (see eq.
(\ref{isovmat})). The $\delta$-dependence of the low lying scissors
becomes quadratic only after inclusion of pairing \cite{Zaw}.

In conclusion we think that above set of citations and argumentations
is convincing enough to state that all the models and methodes
describing the scissors mode without coupling to IVGQR are pure
phenomenological and are therefore of restricted usefulness.

\section{Superdeformation}

As already mentioned, a certain drawback of our approach is that, so
far, we have not included superfluidity into our description.
Nevertheless, our formulas
(\ref{Omeg2fin}, \ref{scimat}, \ref{M1iv}) can be successfully used
for the description of the superdeformed nuclei, where the pairing
is very weak \cite{Lo2000}. For example, applying them to the
superdeformed nucleus $^{152}$Dy ($\delta\simeq 0.6,\,
\hbar\omega_0=41/A^{1/3} \rm{MeV}$), we get
$$E_{iv}=23.6\, {\rm MeV},\quad\quad B(M1)_{iv}=15.9\, \mu_N^2$$
for the isovector GQR and
$$E_{sc}=5.4\, {\rm MeV},\quad\quad B(M1)_{sc}=20.0\, \mu_N^2$$
for the scissors mode. There are not so many results of other
calculations to compare with. As a matter of fact, there are only two
papers considering this problem.

The phenomenological TRM model predicts \cite{Lo2000}:
$$E_{iv}\simeq26\, {\rm MeV},\quad B(M1)_{iv}\simeq26\, \mu_N^2,\quad
E_{sc}\simeq6.1\, {\rm MeV},\quad B(M1)_{sc}\simeq22\, \mu_N^2.$$
The only existing microscopic calculation \cite{Hamam}
in the frame of QRPA with separable forces gives slightly more
information:
$$E_{iv}\simeq28\, {\rm MeV},\quad B(M1)_{iv}\simeq37\, \mu_N^2,$$
$$E_{sc}\simeq5-6\, {\rm MeV},\quad B(M1)_{1^+}\simeq23\, \mu_N^2,
\quad B(M1)_{sc}\simeq0.4\, \mu_N^2.$$ Here $B(M1)_{1^+}$ denotes
the total $M1$ orbital strength carried by the
calculated $K^{\pi}=1^+$ QRPA excitations modes in the energy region
below 20 MeV. The $B(M1)_{sc}$ denotes ``the calculated overlap
probabilities of the QRPA solutions with the synthetic orbital
scissors mode which is defined as
$$|R>=N^{-1}(\L_{1-1}^{\rm n}-q\L_{1-1}^{\rm p})|g.s.>,$$
where $N$ is a normalisation factor. The parameter $q$ is
determined by the requirement that the mode $|R>$ is orthogonal to
the spurious state $|S>\sim \L_{1-1}|g.s.>$'' \cite{Hamam}.

It is easy to see that in the case of IVGQR one can speak, at least,
about qualitative agreement. Our
results for $E_{sc}$ and $B(M1)_{sc}$ are in good agreement with
that of phenomenological model and with $E_{sc}$ and $B(M1)_{1^+}$ of
Hamamoto and Nazarewicz. The very small value of $B(M1)_{syn}$ is
explained quite naturally by the fact, that the synthetic mode does
not treat properly two main ingredients of the scissors mode: Fermi
surface deformation and coupling with IVGQR.

\begin{figure}
\begin{center}
\epsfig{file=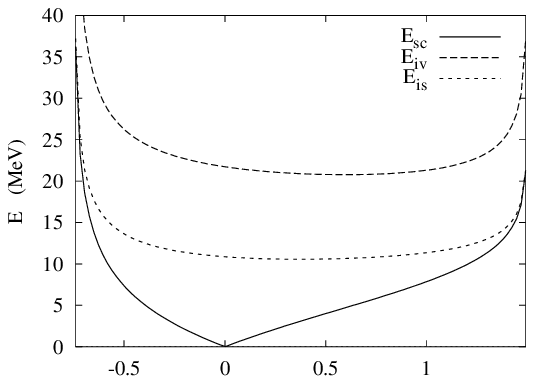,width=12cm}\\
\epsfig{file=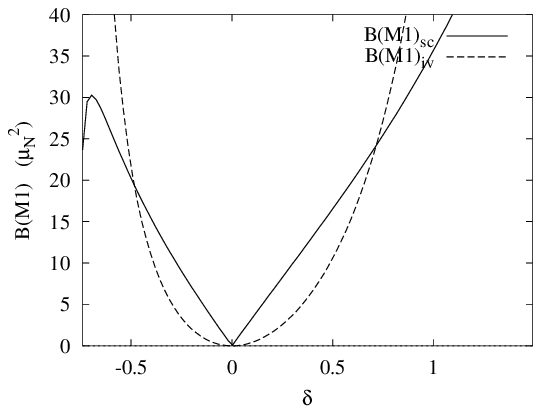,width=12cm}
\end{center}
\caption{Dependences on deformation of energies and $B(M1)$-values
of scissors mode, IVGQR and ISGQR.\label{fig3}}
\end{figure}

Examples of $\delta$-dependences of energies and B(M1)-factors
are shown in Fig.\ref{fig3}.

\section{Conclusion}

In this work we again have considered the issue of the physics behind
the nuclear scissors mode. In spite of 25 years of research and many
valuable contributions to this subject the subtleties of
the scissors mode are still under debate, and in our opinion erroneous
interpretations of the subject continue to appear in the literature.
Surprisingly, no systematic study of the mode in the Bohr-Mottelson
Hamiltonian has yet been carried out, and our purpose here was to fill
that gap. The Bohr-Mottelson Hamiltonian consist of
a harmonic single-particle potential together with a separable Q-Q
interaction. The Q-Q forces have different couplings in the isoscalar
and isovector channels. The isoscalar coupling strength is determined
from Bohr and Mottelson self-consistency condition, leaving the
isovector strength as a free parameter. We adjust it from the fact
that the isovector giant quadrupole resonance
is experimentally known to lie practically at twice the energy of the
isoscalar giant quadrupole mode. With this our model is entirely
fixed and its
solution in the small amplitude limit can be found analytically for
excitation energies and transition amplitudes.

The physics becomes particularly transparent once the TDHF equations
are written down in phase space and the so called Wigner function
moments are introduced. This approach allows one to
establish the optimum set of macroscopic variables:
quadrupole moment, angular momentum, pressure tensor, etc. These
variables are, in
the scheme of our formalism, absolutely unambiguous and, together with
the analytic solution, they allow for a maximum of physical insight.
 At least, the inevitable coupling of the scissors mode with the
isovector giant quadrupole resonance becomes obvious immediately,
already at
the stage of the formulation of the model. Furthermore, the
Fermi surface deformation, whose decisive role in the physics of
the scissors mode is difficult to predict employing naive
phenomenological models, appears in our approach quite naturally.

The eigenvalue equation in the isovector channel yields two
frequencies which are given by
$\Omega_{\pm}=2\bar\omega\sqrt{(1+\delta/3)\pm\sqrt{(1+\delta/3)^2-
\frac{3}{4}\delta^2}}$.
They are distributed in a non symmetric way around twice the harmonic
oscillator frequency $\bar\omega$ and they gradually approach one
another with increasing $\delta$. The low lying frequency corresponds
to the one of the scissors mode proper, whereas the other is the
so-called high lying scissors mode. Our analysis shows that, indeed,
the motion of both modes is "scissors"-like in the sense that the long
symmetry axes of proton and neutron distributions get tilted by a
small angle during their oscillatory motion. Nevertheless both modes
are quite distinct what is revealed by looking at the respective flow
patterns (Figs. \ref{fig1},\ref{fig2}).
The flow lines of the scissors mode are closed ellipses (i.e. mostly
rotational flow) leading to a compression of the long axis, while the
ones of the high lying mode are open hyperbolas (i.e. mostly
irrotational flow) leading to an elongation of the long axis.
The frequency of the scissors mode as a function of mass number turns
out to be about 15\% too low, compared with the experimental data. In
respect to the somewhat crude model we have been employing this may
seem a reasonable agreement. One should, however, mention that we
completely disregarded pairing in our work.
It is generally known that superfluidity
makes the mass parameters smaller, i.e. the frequencies of the modes
become higher. It will be a further task to study whether this
accounts for the missing 15\% in energy.

Other quantities we studied in our model are transition probabilities,
for instance with respect to their deformation dependence. Though for
small deformation most of our results have already been found by other
authors \cite{Zaw} with different methods, we make a point here in
predicting the behaviour for superdeformed nuclei.

We also want to attract the attention to the potential
richness of the set of our equations (\ref{quadr}).
In a further study one may employ
them for the description of the joint dynamics of the isoscalar
and isovector giant monopole and quadrupole resonances plus the
scissors mode in deformed rotating nuclei, the amplitudes of
vibrations being not necessarily small. A large
amplitude motion was already treated in the frame of this approach
to describe the multiphonon giant quadrupole and monopole resonances
\cite{BaSc}. What about two-phonon scissors? The question is not only
academic - the first attempt to interpret some numerical results
as the multiphonon scissors is already known \cite{Sun}.

One may also think to take into account the spin degrees of freedom -
only the number of dynamic equations must be doubled (spin
projections up and down). Then, the theory becomes capable of
describing
spin-flip excitations. As a result, there appears a possibility
of considering the orbital and spin components of the scissors mode
simultaneously.

 It is worth noticing that the set of equations (\ref{quadr}) is
written in the laboratory coordinate system. It allows one to get rid
of any troubles connected with  spurious rotation, because the
total (i.e., isoscalar) angular momentum is
conserved (the last equation of (\ref{isos})). Moreover, the total
angular momentum enters into the dynamic equations - hence one can
study the
behaviour of all modes in rotating nuclei. This would be especially
interesting for the scissors mode in superdeformed (SD) nuclei,
because "The SD bands in nuclei around $^{152}$Dy and $^{192}$Hg are
observed at high spins" \cite{Hamam}. The first interesting results
of calculation interpreted as the rotational band built on the
scissors excitation appeared in \cite{Sun}. The above mentioned
problems shall be investigated in future work.

\section*{Acknowledgments}

We greatfully acknowledge useful comments and discussions with
J.-F. Berger,\\ G.F. Bertsch, M. Girod, H. Goutte, R.V. Jolos,
V.A. Kuz'min, J. Libert, M. Urban.

\section*{Appendix}

It is known that the deformed harmonic oscillator Hamiltonian can be
obtained in a Hartree approximation "by making the assumption that the
isoscalar part of the Q-Q force builds the one-body container well"
\cite{Hilt92}. In our case it is obtained quite easy by summing
formulas (\ref{poten}) and (\ref{noten}):
\begin{equation}
\label{potensc}
V(\br,t)=\frac{1}{2}(V^{\rm p}(\br,t)+V^{\rm n}(\br,t))
=\frac{1}{2}m\,\omega^2r^2+
\kappa_0\sum_{\mu=-2}^{2}(-1)^{\mu}
 Q_{2\mu}(t)q_{2-\mu}(\br).
\end{equation}
In the state of equilibrium (i.e. in the absence of external field)
$Q_{2\pm1}=Q_{2\pm2}=0$. Using the definition \cite{BM}
$Q_{20}=Q_{00}\frac{4}{3}\delta$ and the formula
$q_{20}=2z^2-x^2-y^2$ we obtain the potential of the
anisotropic harmonic oscillator
$$V(\br)=\frac{m}{2}[\omega_x^2(x^2+y^2)+\omega_z^2z^2]$$
with oscillator frequencies
$$\omega_x^2=\omega_y^2=\omega^2(1+\sigma\delta), \quad
\omega_z^2=\omega^2(1-2\sigma\delta),$$
where $\di \sigma=-\kappa_0\frac{8Q_{00}}{3m\omega^2}$.
The definition of deformation parameter $\delta$ must be reproduced
by the harmonic oscillator wave functions, that allows one to fix the
value of $\sigma$. We have:
$$Q_{00}=\frac{\hbar}{m}(\frac{\Sigma_x}{\omega_x}
+\frac{\Sigma_y}{\omega_y}+\frac{\Sigma_z}{\omega_z}),\quad
Q_{20}=2\frac{\hbar}{m}(\frac{\Sigma_z}{\omega_z}
-\frac{\Sigma_x}{\omega_x}),$$
where $\di \Sigma_x=\Sigma_{i=1}^A(n_x+\frac{1}{2})_i$ and $n_x$
is the oscillator quantum number. Using the self consistency condition
$$\Sigma_x\omega_x=\Sigma_y\omega_y
=\Sigma_z\omega_z=\Sigma_0\omega_0,$$
where $\Sigma_0$ and $\omega_0$ are defined for spherical nucleus,
we get
$$\frac{Q_{20}}{Q_{00}}=2\frac{\omega_x^2-\omega_z^2}
{\omega_x^2+2\omega_z^2}=\frac{2\sigma\delta}{1-\sigma\delta}
=\frac{4}{3}\delta.$$
Solving last relation with respect of $\sigma$ we find
\begin{equation}
\label{sigma}
\sigma=\frac{2}{3+2\delta}.
\end{equation}
Now we can write final expressions for oscillator frequences
$$\omega_x^2=\omega_y^2=\omega^2\frac{1+\frac{4}{3}\delta}
{1+\frac{2}{3}\delta}, \quad
\omega_z^2=\omega^2\frac{1-\frac{2}{3}\delta}
{1+\frac{2}{3}\delta}$$
and the self consistent value of the strength constant
$$\kappa_0=-\frac{m\omega^2}{4Q_{00}}\frac{1}{1+\frac{2}{3}\delta}.$$
The condition for volume conservation
$\omega_x\omega_y\omega_z=const=\omega_0^3$ makes $\omega$
$\delta$-dependent:
$$\omega^2=\omega_0^2\frac{1+\frac{2}{3}\delta}
{(1+\frac{4}{3}\delta)^{2/3}(1-\frac{2}{3}\delta)^{1/3}}.$$
$Q_{00}$ depends on $\delta$ as
$$Q_{00}=Q_{00}^0\frac{1}
{(1+\frac{4}{3}\delta)^{1/3}(1-\frac{2}{3}\delta)^{2/3}},$$
where $Q_{00}^0=A\frac{3}{5}R^2,\,R=r_0A^{1/3}.$

\end{document}